\documentclass{PoSi}

\usepackage{multicol}
\usepackage{cite}
\usepackage{overpic}
\usepackage{rviewport}
\usepackage{xspace}
\usepackage{xcolor}
\usepackage{tabularx}
\usepackage{booktabs}

\def\EeV{\ifmmode {\mathrm{Ee\kern -0.07em V}}\else
                   \textrm{Ee\kern -0.07em V}\fi\xspace}
\def\TeV{\ifmmode {\mathrm{Te\kern -0.07em V}}\else
                   \textrm{Te\kern -0.07em V}\fi\xspace}
\def\GeV{\ifmmode {\mathrm{Ge\kern -0.07em V}}\else
                   \textrm{Ge\kern -0.07em V}\fi\xspace}
\def\eV{\ifmmode {\mathrm{\ e\kern -0.07em V}}\else
                   \textrm{e\kern -0.07em V}\fi\xspace}
\def\gcm{\ensuremath{\mathrm{g/cm}^2}\xspace}
\def\Xmax{\ifmmode {X_\mathrm{max}}\else
                   {$X_\mathrm{max}$}\fi\xspace}%
\def\Xmumax{\ifmmode {X_\mathrm{max}^\mu}\else
                   {$X_\mathrm{max}^\mu$}\fi\xspace}%
\def\sigmaXmax{\ifmmode {\sigma(X_\mathrm{max})}\else
                   {$\sigma(X_\mathrm{max})$}\fi\xspace}%
\def\meanXmax{\ifmmode {\langle X_\mathrm{max}\rangle}\else
                   {$\langle X_\mathrm{max}\rangle$}\fi\xspace}%
\def\meanLnA{\ifmmode {\langle \ln A\rangle}\else
                   {$\langle \ln A\rangle$}\fi\xspace}%

\newcommand{\energy}[1]{\ensuremath{10^{#1}}\,\eV}
\newcommand{\Energy}[2]{\ensuremath{#1 \times 10^{#2}}\,\eV}
\def\SwiftBAT{\emph{Swift}-BAT\xspace}
\def\FermiLAT{\emph{Fermi}-LAT\xspace}

\graphicspath{{figs/}}
\DeclareGraphicsExtensions{.pdf,.png,.jpg}

\pdfpageattr{/Group << /S /Transparency /I true /CS /DeviceRGB>>}

\title{Highlights from the Pierre Auger Observatory}

\ShortTitle{Auger Highlights}

\author{\speaker{Michael Unger}$^a$ for the Pierre Auger Collaboration$^b$\\
\llap{$^a$}Karlsruhe Institute of Technology, Institut f\"ur Kernphysik,
           Karlsruhe, Germany\\
\llap{$^b$}Observatorio Pierre Auger, Av.\ San Mart\'in Norte 304, 5613 Malarg\"ue,
           Argentina\\
E-mail: \href{mailto:auger_spokespersons@fnal.gov}
             {\rm auger\_spokespersons@fnal.gov}\\
Full author list: \href{http://www.auger.org/archive/authors_icrc_2017.html}
                  {\rm http://www.auger.org/archive/authors\_icrc\_2017.html}}

\abstract{ In this contribution we summarize the highlights from the
  Pierre Auger Observatory presented at the 35th International Cosmic
  Ray Conference.  We discuss the update of the measurement of the
  energy spectrum of cosmic rays over a wide range of energies
  ($10^{17.5}$ to above $10^{20}$\,eV), studies of the cosmic-ray mass
  composition with the fluorescence and surface detector
  of the Observatory, the discovery of a large-scale anisotropy in the
  arrival direction of cosmic rays above \Energy{8}{18} and
  indications of anisotropy at intermediate angular scales above
  \Energy{4}{19}. Moreover, we report on tests of hadronic
  interactions beyond LHC energies, multi-messenger analyses with
  neutral primaries and the progress of the upgrade of the
  Observatory, AugerPrime, aimed at elucidating the origin of the
  observed flux suppression at ultra-high energies.}

\FullConference{35th International Cosmic Ray Conference --- ICRC2017\\
                10--20 July, 2017\\
                Bexco, Busan, Korea}

\let\OLDthebibliography\thebibliography
\renewcommand\thebibliography[1]{
  \OLDthebibliography{#1}
  \setlength{\parskip}{0pt}
  \setlength{\itemsep}{0pt plus 0.3ex}
  \setlength{\columnsep}{-20cm}
}
\begin{document}

\section{Introduction}
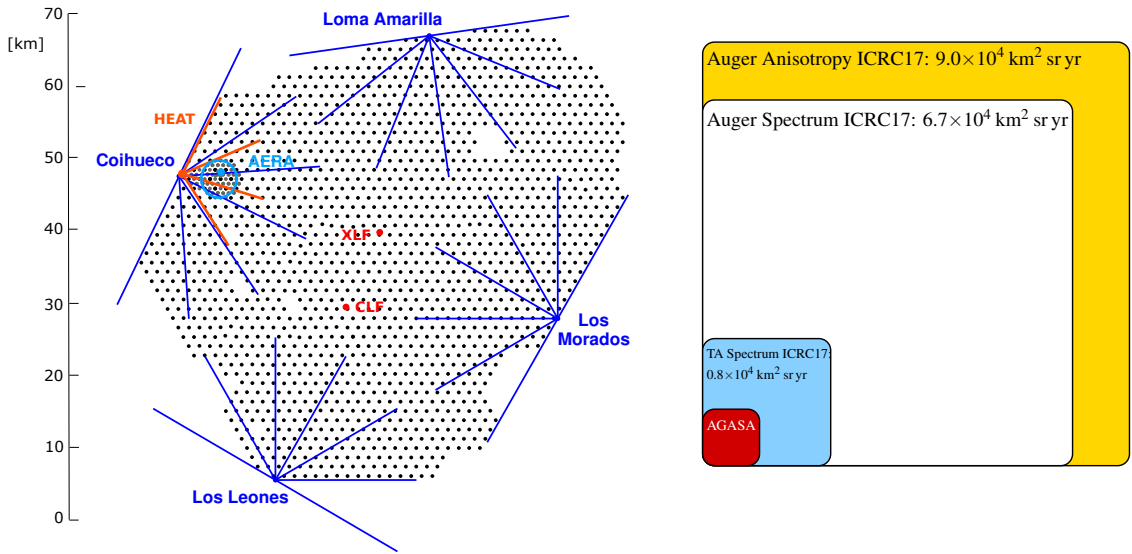
\begin{figure}[t]
\centering
\includegraphics[width=0.54\textwidth]{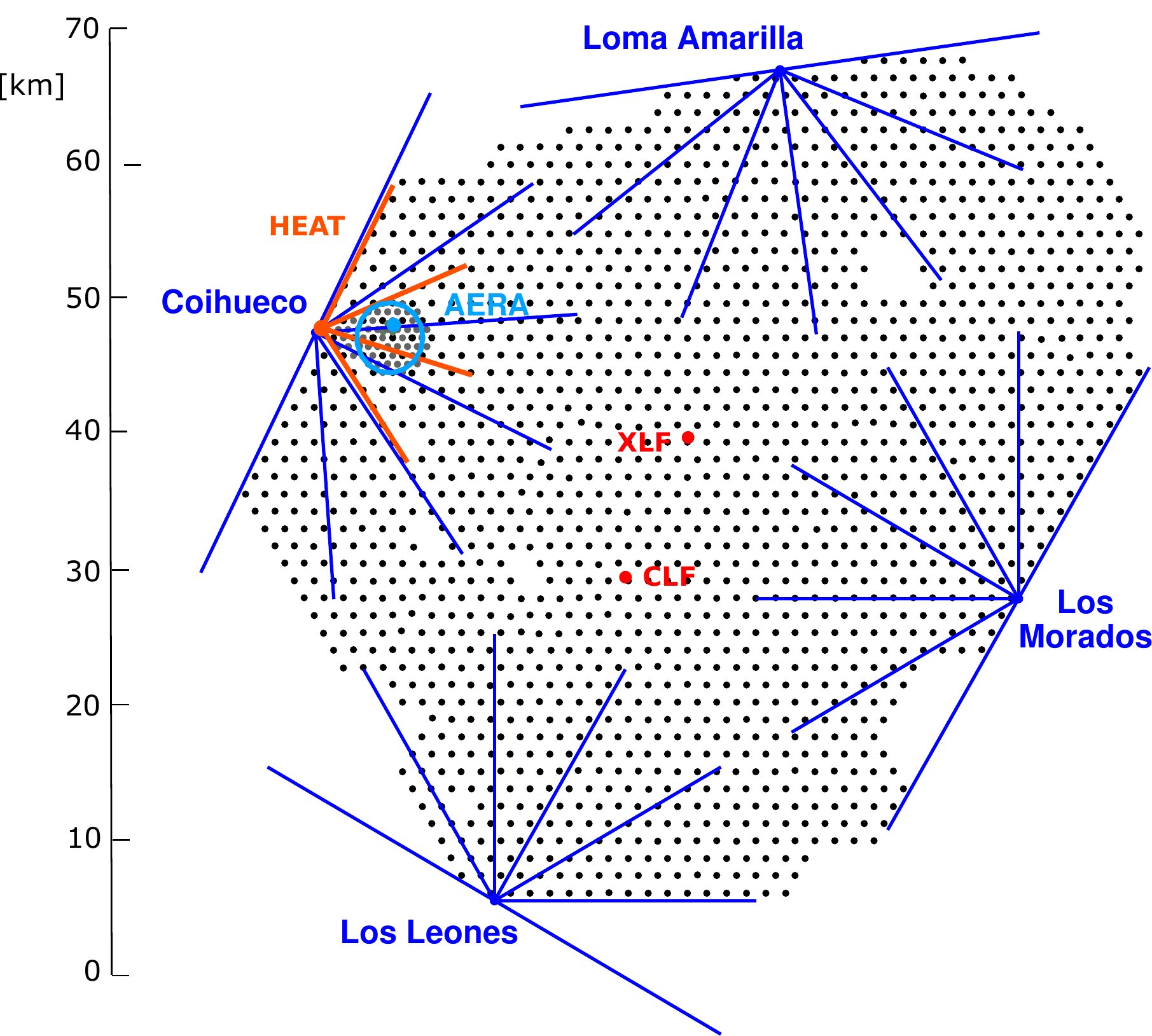}
\begin{overpic}[clip, rviewport=-0.15 -0.2 1 1.05,width=0.43\textwidth]{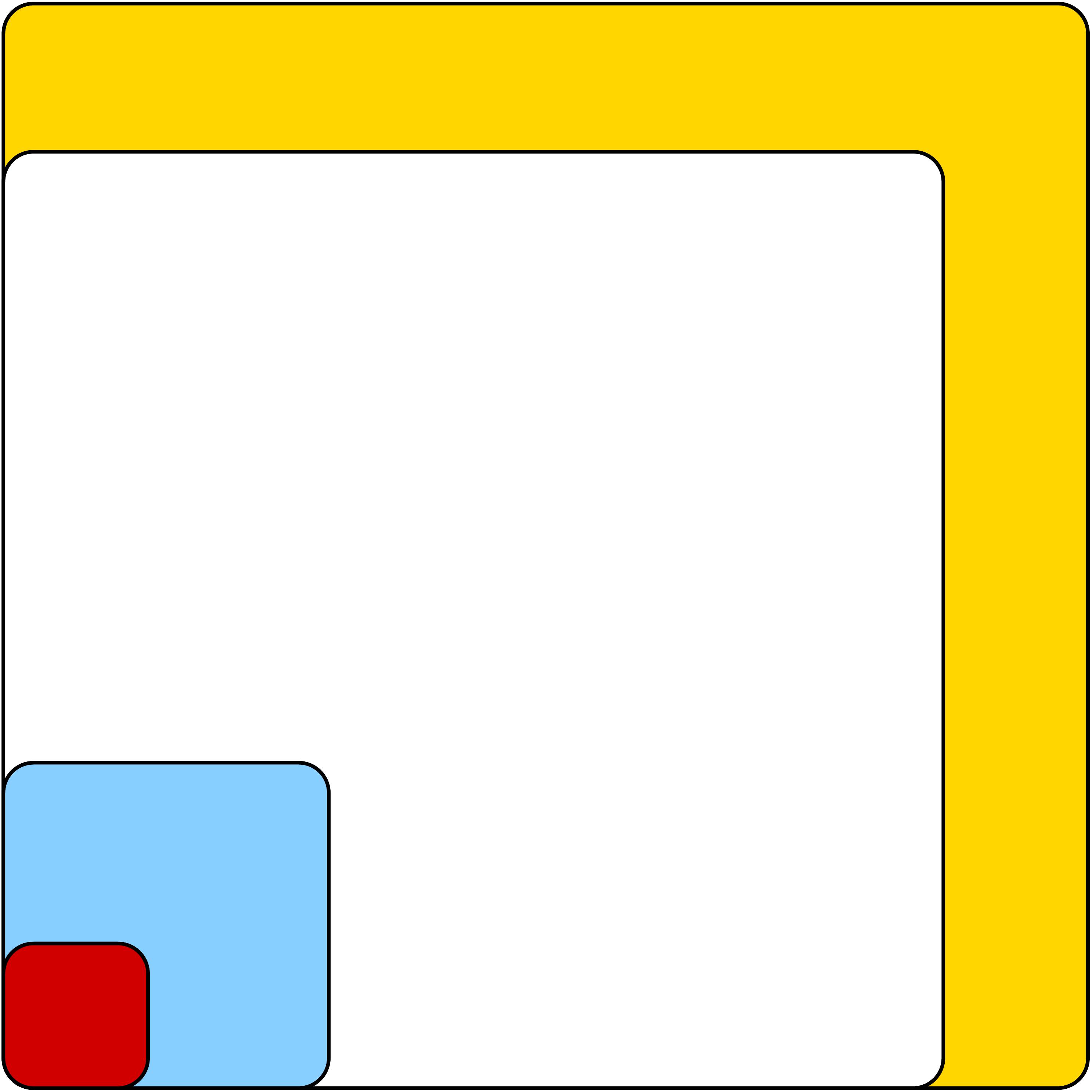}
\put (13,91.5) {\scriptsize \scalebox{.94}{Auger Anisotropy ICRC17: 9.0$\times$10$^4$\,km$^2$\,sr\,yr}}
\put (13,80) {\scriptsize \scalebox{0.94}{Auger Spectrum ICRC17: 6.7$\times$10$^4$\,km$^2$\,sr\,yr}}
\put (13,36.5) {\tiny \scalebox{0.85}{TA Spectrum ICRC17:}}
\put (13,32.5) {\tiny \scalebox{0.85}{0.8$\times$10$^4$\,km$^2$\,sr\,yr}}
\put (13,23) {\tiny \scalebox{.9}{\color{white} AGASA}}
\end{overpic}
\caption{{\itshape Left:} Layout of the Pierre Auger
  Observatory. Water-Cherenkov detectors are shown as black dots and
  the azimuthal field of views of the 27 fluorescence telescopes is
  indicated by blue and red lines. The location of the two laser
  facilities (CLF and XLF) for the monitoring of the aerosol content
  in the atmosphere are shown with red dots and the area equipped with
  radio antennas (AERA) is marked with a light-blue circle.
  {\itshape Right:} Illustration of the exposures available for two of
  the Auger analyses described in this paper (yellow and white) and
  the exposures collected by the Telescope Array (blue) and AGASA
  (red).}
\label{fig:exposure}
\end{figure}
In this article we summarize highlights from the presentations of the
Pierre Auger Coll. at the 35th International Cosmic Ray
Conference~\cite{contributions, Aab:2017njo}.  The Pierre Auger
Observatory~\cite{ThePierreAuger:2015rma} is the largest facility
built so far to detect cosmic rays.  It is located in the province of
Mendoza, Argentina and has been in operation since 2004.  Cosmic rays
are studied by combining a Surface Detector (SD) and a Fluorescence
Detector (FD) to measure extensive air showers. The SD consists of
1600 water-Cherenkov detectors on a 1500\,m triangular grid (SD\,1500)
over an area of $\sim$3000~km$^2$, and of an additional 61 detectors
covering 23.5~km$^2$ on a 750 m grid (SD\,750).
The SD
is overlooked by 27 fluorescence telescopes located in five buildings
on its periphery. 24 telescopes cover 30$^\circ$ in azimuth and
elevations from $1.5^\circ$ to 30$^\circ$ above the horizon.  Three
additional telescopes enlarge the elevation coverage to 60$^\circ$ for
low-energy showers.  The SD is used to measure photons and charged
particles at ground level with a duty cycle near 100\% and the FD to
observe the longitudinal development of air showers in the atmosphere
during dark nights, and under favorable meteorological conditions with
a duty cycle of about 13\%.  This setup is complemented by the Auger
Engineering Radio Array (AERA) to study radio emission from air
showers~\cite{aera1}.  The layout of the Observatory is shown
in the left panel of Fig.~\ref{fig:exposure} and the exposure
collected between January 2004 and December 2016 is illustrated in the
right panel. The exposure thus far collected is close to
$10^5$\,km$^2$\,sr\,yr for studies of anisotropies in the arrival
directions of cosmic rays.  The large exposure and an 85\% coverage of
the celestial sphere are the prerequisites to perform the precise
measurements of the flux, composition and anisotropy of
ultra-high-energy cosmic rays (UHECRs) described in the following.

\begin{figure}[t]
\centering
\includegraphics[clip,rviewport=0 -0.12 1.05 1,width=0.37\textwidth]{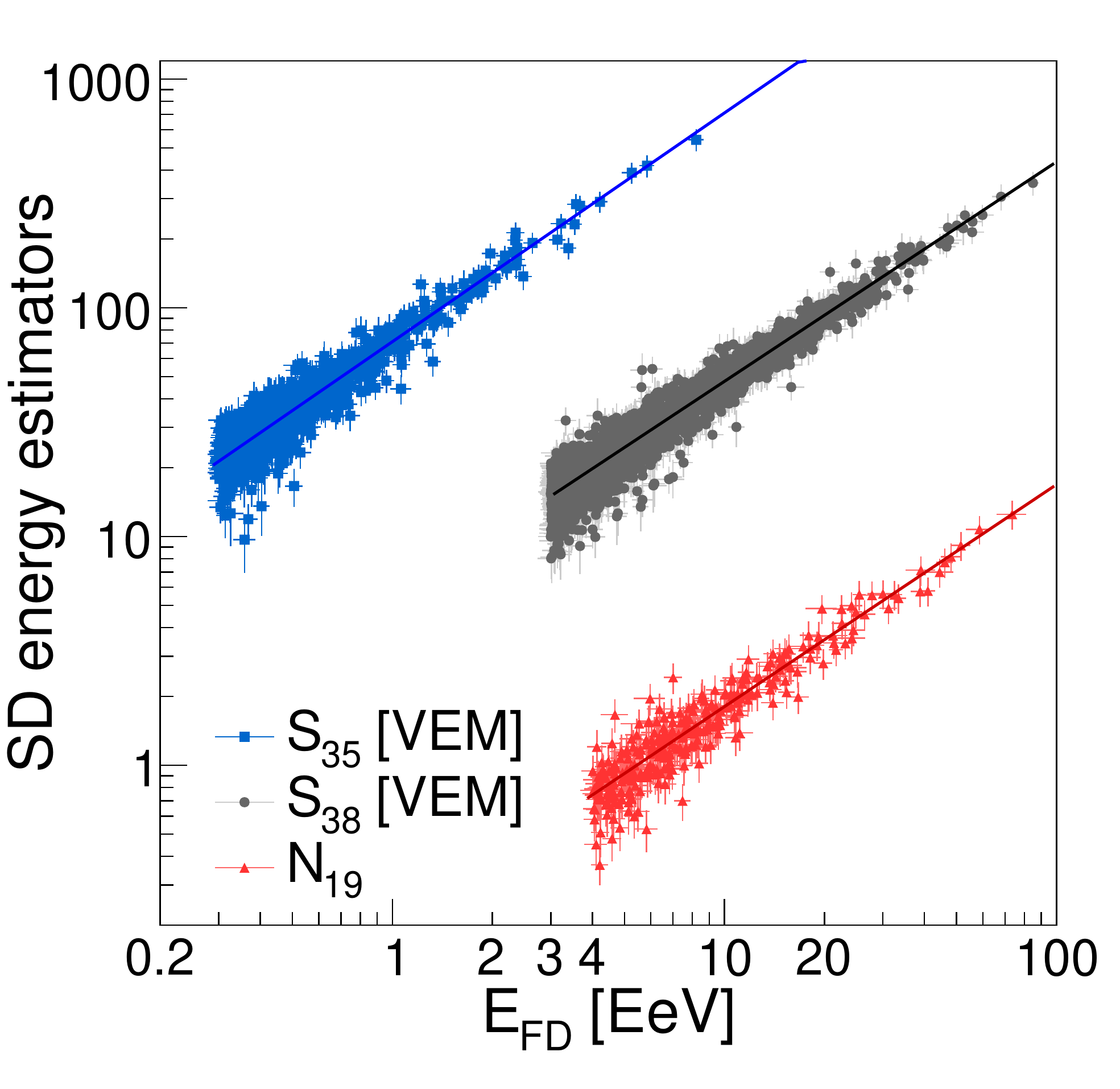}\hfill
\begin{overpic}[width=0.58\textwidth]{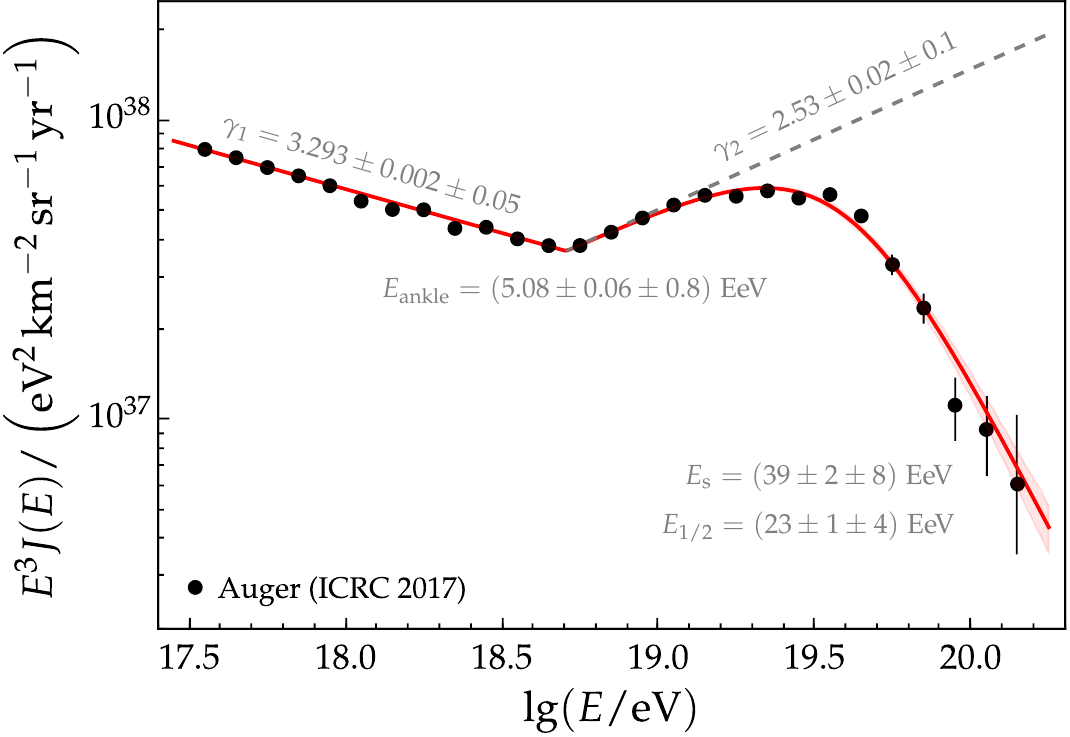}
\end{overpic}
\caption{{\itshape Left}: Energy calibration of the surface
  detector. The shower size measured for near-vertical events with the
  SD\,1500 ($S_{38}$) and SD\,750 ($S_{35}$) array and for inclined
  showers ($N_{19}$) is shown as a function of the energy measured
  with the fluorescence telescopes ($E_{\rm FD}$).  {\itshape Right}:
  Combined energy spectrum. The red line shows a fit to the spectrum
  with a broken power law and a suppression at ultra-high
  energies. The gray dashed line indicates the same broken power law
  without suppression. The fitted spectral indices and energies of the
  break and suppression are superimposed together with their statistical and
  systematic uncertainties.}
\label{fig:spectrum}
\end{figure}

\section{Energy Spectrum}
\label{sec:spec}
We use four different data sets to derive the energy spectrum of
cosmic rays~\cite{spectrum}.  The two ``near-vertical'' data samples
consist of showers arriving at zenith angles $\leq 60^\circ$ detected
by the SD\,1500 and SD\,750 arrays with an energy threshold of
\Energy{3}{18} and \Energy{3}{17}, respectively. ``Inclined'' showers
with zenith angles between 60$^\circ$ and 80$^\circ$ are studied with
the SD\,1500 array above \Energy{4}{18}. Finally, the ``hybrid''
sample consists of events above \energy{18} detected by the FD
simultaneously with at least one station of the SD. From each of the
data sets an independent energy spectrum is derived and, after
establishing the compatibility between the individual spectra, the
four measurements are statistically combined to obtain our best
estimate of the energy spectrum of cosmic rays above \Energy{3}{17},
from a total exposure of 6.7$\times$10$^4$\,km$^2$\,sr\,yr, collected
between January 2004 and December 2016.

The method to derive the spectra is entirely data-driven and
free of model-dependent assumptions about hadronic interactions in air
showers. For the surface detector data, we transform the
measured shower sizes to a size estimator that is independent of
zenith angle by using the method of constant intensities for the
near-vertical data~\cite{Hersil:1961zz} and templates of the footprint
of the particle densities at the ground for the inclined data
set~\cite{Aab:2014gua}. These attenuation-corrected shower sizes
are used as energy estimates after calibrating them
with the calorimetric energy available for events that
have been observed simultaneously with the surface and fluorescence
detectors, as shown in the left panel of Fig.~\ref{fig:spectrum}.

The main ingredients for the energy scale of the Observatory are a
precise laboratory measurement of the fluorescence
yield~\cite{Ave:2012ifa}, the optical calibration of the fluorescence
telescopes~\cite{Brack:2013bta} and the monitoring of the light
attenuation in the atmosphere~\cite{Abreu:2013qtw}. At this conference
we presented improvements of the data-driven estimate of the
``invisible energy'' (i.e.\ the amount of energy in air showers
carried away by muons and neutrinos), the event reconstruction at low
energies, the spectral calibration of the telescopes and the
determination of the aerosol content of the atmosphere~\cite{spectrum,
  aerosolupdate, Aab:2017ctu}. These improvements result in an update
of the energy scale of the Observatory with changes of $\leq 4\%$. The
overall systematic uncertainty of the energy scale remains at
14\%~\cite{energyscaleuncert}. An independent determination of the
Auger energy scale using the radiation energy in the radio signal of
air showers with AERA might be possible in the near
future~\cite{aera2, Aab:2016eeq}.

The combined energy spectrum is shown in the right panel of
Fig.~\ref{fig:spectrum}. To quantify the features of the spectrum, we
fitted the flux with a power law allowing for a break in the spectral
index at $E_{\rm ankle}$ and a suppression of the flux at ultra-high
energies $\displaystyle \varpropto \left(1+(E/E_{\rm s})^{\Delta_{\rm
    s}}\right)^{-1}$.  As can be seen, we find that the position of
the ``ankle'' in the cosmic-ray spectrum is at $E_{\rm ankle} =
(5.08\pm 0.06 ({\rm stat.})  \pm 0.8 ({\rm syst.}))~\EeV$ where the
spectral index hardens by $\Delta\gamma \sim - 0.76$. A power-law
extension of the flux above the ankle towards ultra-high energies is
clearly excluded by our measurement and we find a suppression energy
of $E_{\rm s} = (39\pm 2 ({\rm stat.})  \pm 8 ({\rm syst.}))~\EeV$ and
a softening of $\Delta_{\rm s} = (2.5\pm 0.1 ({\rm stat.})  \pm 0.4
({\rm syst.}))$. The current exposure allows us to probe the spectrum
up to \energy{20.15} where we find that the flux is suppressed by more
than an order of magnitude with respect to a power-law
continuation. The energy at which the integral flux drops by a factor
of two below what would be expected without suppression is found to be
$E_{1/2} = (23\pm 1 ({\rm stat.})  \pm 4 ({\rm syst.}))~\EeV$. This
value is at considerable odds with $E_{1/2} = 53~\EeV$ as
predicted in~\cite{Berezinsky:2002nc} for the suppression of the flux of
protons produced uniformly in extragalactic space suffering energy
losses due to photo-pion production in interactions with the photons
of the cosmic microwave background radiation during propagation to
Earth.

\section{Mass Composition}
\label{sec:mass}
\begin{figure}[t]
\centering
\includegraphics[clip,rviewport=0 -0.12 1.1 1,width=0.36\textwidth]{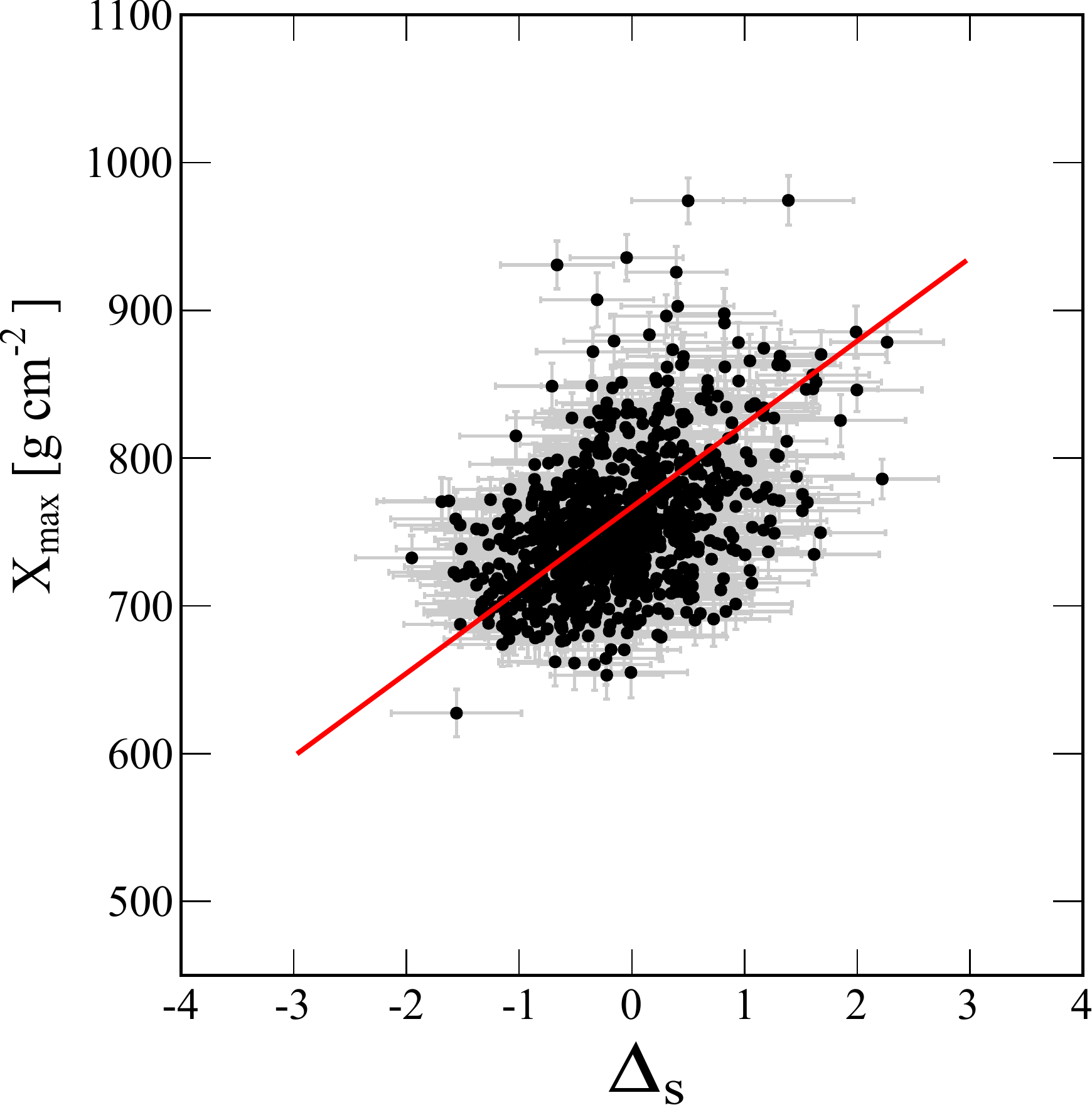}
\begin{overpic}[clip,rviewport=0.02 0.05 0.955 1,width=0.63\textwidth]{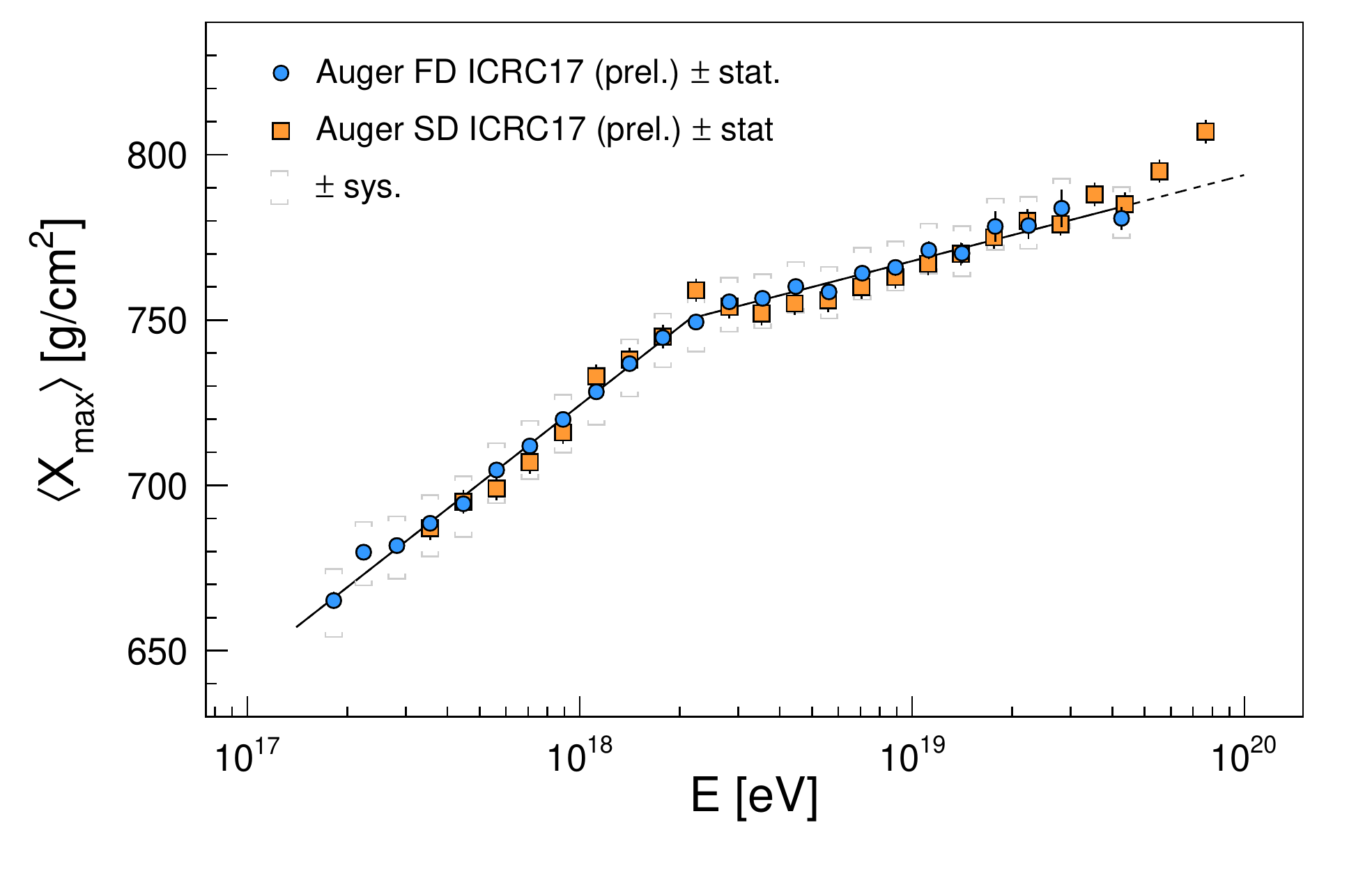}
\put (19,25) {\color{gray}\scriptsize \rotatebox{39}{\scalebox{.95}{$(79\pm 1)~\gcm$/decade}}}
\put (60,31.5) {\color{gray}\scriptsize \rotatebox{16}{\scalebox{.95}{$(26\pm 2)~\gcm$/decade}}}
\end{overpic}
\caption{{\itshape Left}: Calibration of the SD variable
  $\Delta_{\rm S}$ with \Xmax measured with the FD. {\itshape Right}:
  Average shower maximum, \meanXmax, from FD (blue dots) and SD
  (orange squares).  A fit of the FD measurements with a broken linear
  function in $\lg(E)$ is displayed as a solid black line. Its
  extrapolation to high energies is indicated with a dashed line. The
  values of the fitted elongation rates are shown in gray.}
\label{fig:rawXmax}
\end{figure}

The mass composition of cosmic rays is studied by measuring the
atmospheric depth, $\Xmax$, at which the number of particles in an air
shower reaches its maximum. On average light primaries penetrate
deeper into the atmosphere than heavy primaries ($\meanXmax \varpropto
\lg(E/A)$ where $A$ is the nuclear mass number) and the position of
shower maximum fluctuates more for light than heavy primaries.  The
measurement of the average shower maximum as a function of energy with
the FD is shown as blue dots in the right panel of
Fig.~\ref{fig:rawXmax}. This measurement is based on 42,466 events
above \energy{17.2}, out of which 62 have been detected at the highest
energy ($E>\energy{19.5}$)~\cite{XmaxFD}. Biases from detector
effects, e.g.\ due to the limited acceptance of the fluorescence
telescopes, are avoided using the analysis methods explained
in~\cite{Aab:2014kda}.  We find that the \meanXmax evolves with energy
at a rate of $79~\gcm$/decade and $26~\gcm$/decade with a break in the
evolution at \energy{18.33 \pm 0.02}, i.e.\ about a factor 2.5 lower
in energy than the break in flux reported in the previous
section. Predictions of the energy evolution of \meanXmax from air
shower simulations are around $60~\gcm$/decade, irrespective of the
mass of the primary and the model used to simulate hadronic
interactions. It can therefore be concluded that the average mass of
cosmic rays evolves towards a lighter composition between $10^{17.2}$
and \energy{18.33}, qualitatively consistent with a transition from a
heavy Galactic composition, to a light extragalactic composition. At
higher energies the trend is reversed and the average mass increases
with energy.

The statistics collected with the relatively low duty
cycle of FD do not yet allow us to study the composition at energies
where the flux suppression is observed. At these energies we rely
instead on measurements with the SD.  However, previous measurements
from Auger have shown that air shower simulations fail to reproduce
salient features of the SD data~\cite{Aab:2014dua, Aab:2014pza,
  Aab:2016hkv}. We therefore follow a similar approach as for the
energy spectrum and calibrate the mass observable from the SD with
\Xmax from the FD.

Here we present results using the time profiles of
the signals recorded with the SD stations, employing the fact that the
{\itshape risetime}~\cite{Watson:1974tf} of these signals depends on
the distance of the shower maximum to the ground and the relative
amount of muons and electrons detected. For each event, we construct
the risetime-related variable $\Delta_{\rm S}$~\cite{XmaxDelta} and
correlate it with \Xmax from the FD as displayed in the left panel of
Fig.~\ref{fig:rawXmax}.  After the calibration function that relates
$\Delta_{\rm S}$ to \Xmax is determined separately for the SD\,1500
and SD\,750 arrays, the energy evolution of \meanXmax can be measured
with the surface detector over a wide range of energies.  The
comparison of \meanXmax from the FD and SD is shown in the right panel
of Fig.~\ref{fig:rawXmax}. As can be seen, the two measurements are in
good agreement, as is to be expected due to the cross-calibration. At
ultra-high energies the superior statistics of the SD (517 events above
$E>\energy{19.5}$) give two data points more than obtained with the
FD.  Interestingly, these two last bins above \energy{19.6} do not
follow the evolution of the composition seen at lower energies.  This
might be an indication that the increase of the average nuclear mass
with energy is slowing at the highest energies. However, the final
point is just $\sim 3 \sigma$ above the low-energy extrapolation of
the \meanXmax from SD and therefore more data are needed to study this
energy region.

\begin{figure}[t]
\includegraphics[clip, rviewport= 0 0.05 0.96 1,width=0.48\textwidth]{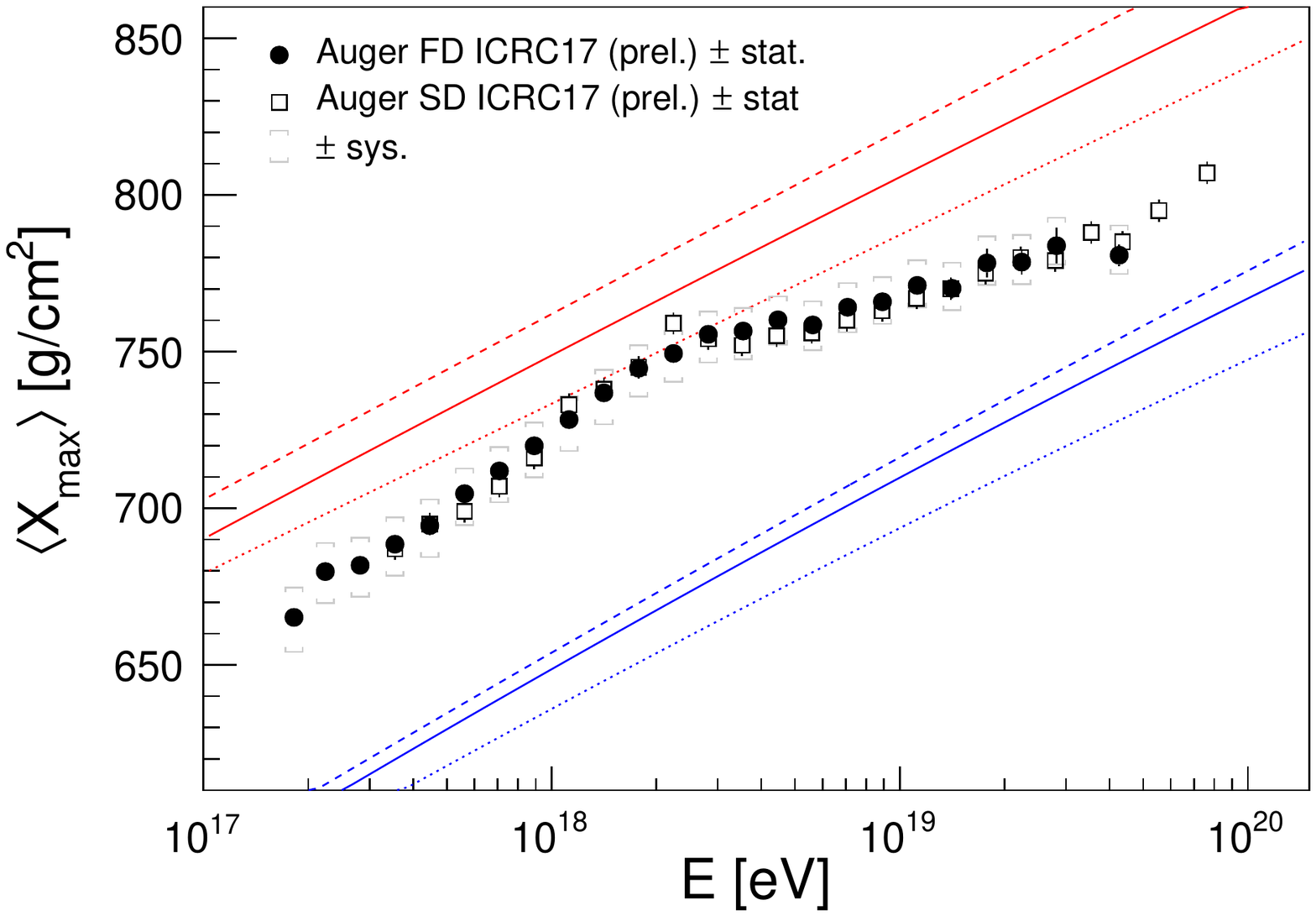}\hfill
\includegraphics[clip, rviewport= 0 0.05 0.96 1,width=0.48\textwidth]{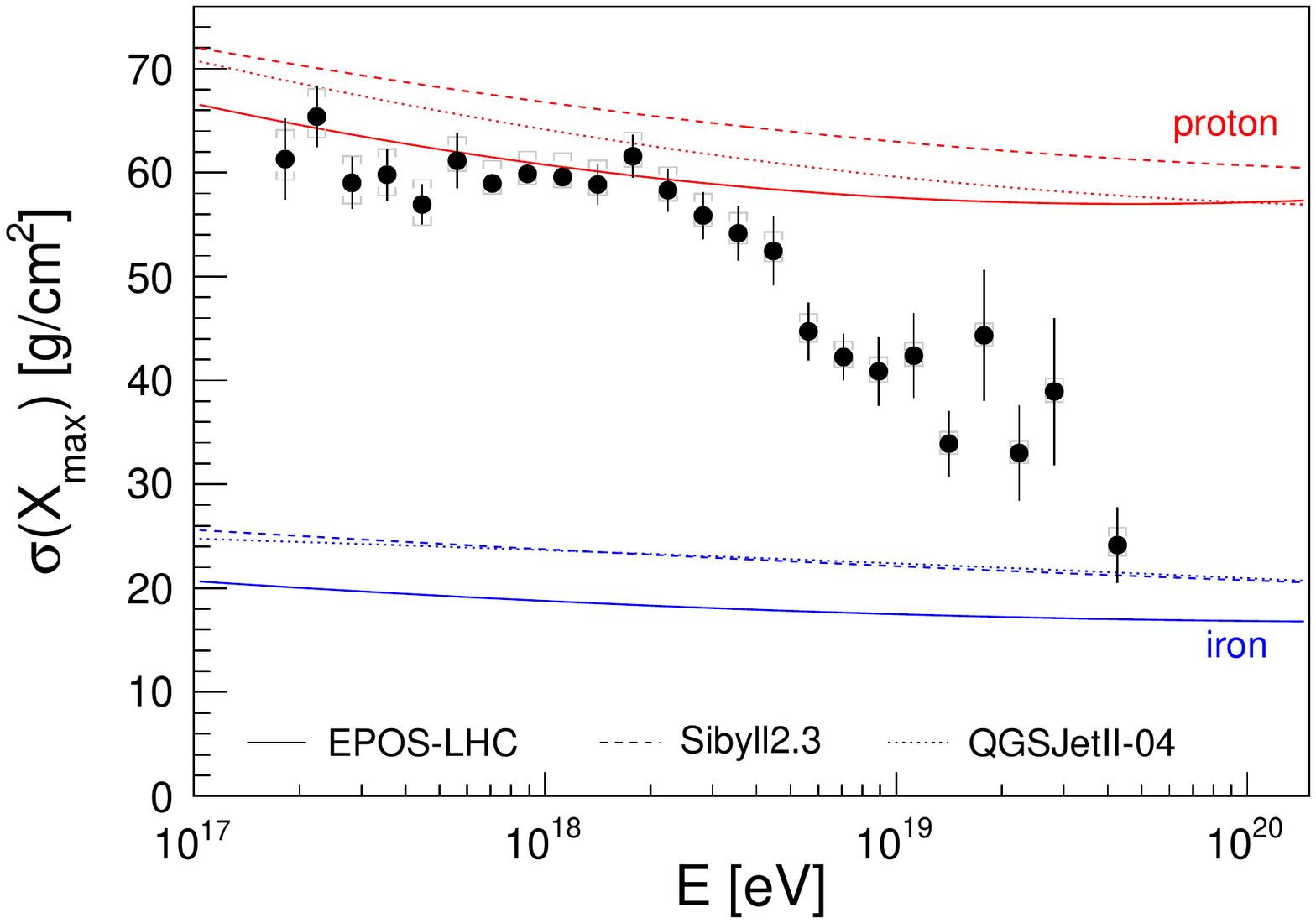}
\caption{Comparison of \meanXmax (left) and \sigmaXmax (right) to predictions
  from simulations of proton- and iron-induced air showers.}
\label{fig:xmax}
\end{figure}

Additional information on the composition can be obtained at the expense
of an increased model dependence by comparing the measured \meanXmax
values directly to predictions of air shower simulations. This is
shown in the left panel of Fig.~\ref{fig:xmax}, where it can be seen
that around the break of \meanXmax at \energy{18.3}, the
composition is very light (even compatible with a pure proton
composition for shower simulations with the {\scshape QGSJetII}
hadronic interaction model) and the average mass increases to a value
between proton and iron at the highest energies, with the exact
relative position depending on the model. Note that the three hadronic
interaction models used for the simulations shown in
Fig.~\ref{fig:xmax} have all been tuned to match data from the
LHC~\cite{Ostapchenko:2010vb, Pierog:2013ria, Riehn:2015oba}. Older
versions of these models are at odds with laboratory measurements at
equivalent fixed-target beam energies of up to \energy{17}, and can
therefore not be used to reliably interpret the data.

In addition to the average of \Xmax, its
standard deviation, \sigmaXmax, can also be determined from the data of the
FD and the results are shown in the right panel of
Fig.~\ref{fig:xmax}. A large value of \sigmaXmax can originate from
either a light composition or a mixed composition, where in the latter
case the difference in \meanXmax of different nuclei adds to the
overall width of the \Xmax distribution. The data at low energies is
compatible with both possibilities. At high energies,
\sigmaXmax decreases indicating a rather pure and heavy
composition. It was shown at this conference~\cite{XmaxFD} that using
the approach of~\cite{Abreu:2013env} one can find compositions that
result in values of \meanXmax and \sigmaXmax that are compatible with
the Auger data if hadronic interactions in air showers are similar
to the ones predicted by {\scshape EposLHC} or {\scshape Sibyll2.3}.
However, for simulations with {\scshape QGSJetII-04}, the derived
average mass is too light to produce shower fluctuations as narrow as
the measured ones and this model is at odds with our data.

\begin{figure}[t]
\centering
\includegraphics[width=0.6\linewidth]{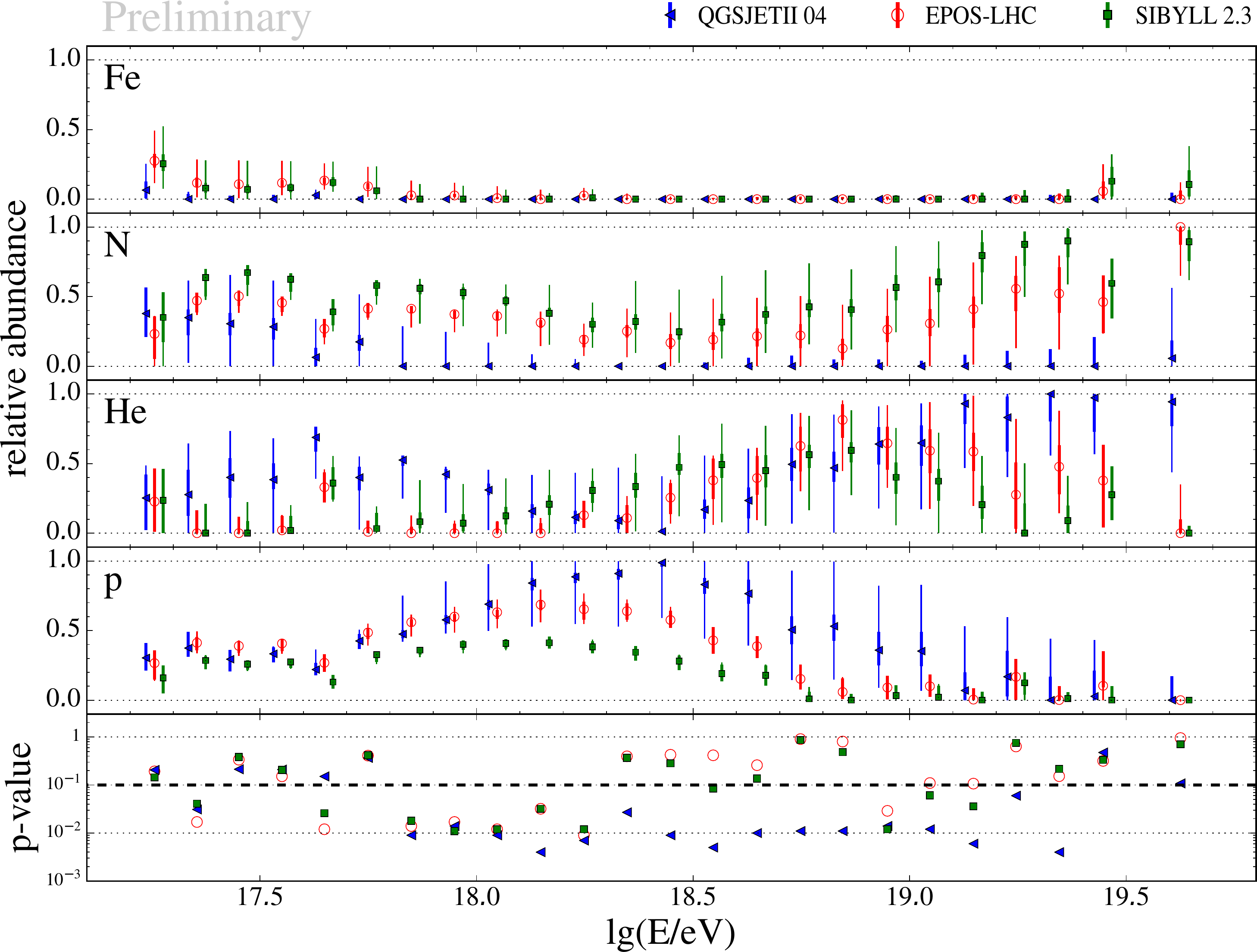}
\caption{Results from a fit of the \Xmax distributions with a
  superposition of p-, He-, N- and Fe-induced air showers. The upper
  four panels show the best-fit mass fractions and the goodness of fit
  is displayed in the lowest panel. Thick error bars denote the statistical
  uncertainties, thin error bars the systematic ones.}
\label{fig:comp}
\end{figure}

An interpretation of the full \Xmax distribution in each energy bin is
achieved by fitting a superposition of \Xmax-templates obtained from
simulations of p-, He-, N- and Fe-induced air showers to the data. At
this conference we presented an update of our previous
study~\cite{Aab:2014aea} with increased statistics at high energies
and for the first time also for data below \energy{17.8}. The
resulting mass fractions are shown in Fig.~\ref{fig:comp}.
At high energies they are compatible with our earlier
finding that the composition is dominated by a single elemental group
starting from protons below the ankle and going through helium to
nitrogen as the energy increases. Depending on the hadronic
interaction model, a small proton fraction may persist up to
ultra-high energies and there might be an iron contribution emerging
above \energy{19.4}. The aforementioned difficulty of describing
\meanXmax and \sigmaXmax with {\scshape QGSJetII-04} is also visible
in the lower panel of Fig.~\ref{fig:comp}, where the probability of the
fits is shown. The fit probabilities obtained with {\scshape QGSJetII-04} are
consistently low at around 0.01 above \energy{17.8}. Therefore, the mass
fractions obtained with {\scshape QGSJetII-04} should be interpreted
with care.

At the lowest energies, we find hints for a contribution from iron
primaries that disappears rapidly with increasing energy. The proton
fraction between $10^{17.2}$ and \energy{17.7} is found to be
approximately constant at a value of 38\%, 28\% and 25\% for {\scshape
  EposLHC}, {\scshape Sibyll2.3} and {\scshape QGSJetII-04}
respectively. These estimates of the proton fraction are based on 7498
events and have a statistical and systematic uncertainty of $2\%$ and
$\leq9\%$ respectively. For comparison, a recent attempt to estimate
the light-mass fraction from $\Xmax$ was based on 118 events
only~\cite{Buitink:2016nkf}. The mass fractions presented here
complement the findings of the KASCADE-Grande Coll. which
reported a ``knee'' in the flux of the iron component at \energy{16.9}
and an ``ankle'' of the light component at
\energy{17.1}~\cite{Apel:2011mi, Apel:2013ura}. It can be concluded
that the new results from the Pierre Auger Observatory on the mass
composition at low energies give important experimental constraints to
the modeling of a possible transition from a heavy Galactic to a light
extragalactic cosmic-ray component between $10^{17}$ and \energy{18}.

We end this section with the usual caveats about the model dependence
of the interpretation of air-shower observables in terms of mass. The
mass fractions derived from the \Xmax distribution are very sensitive
to details of the modeling of hadronic interactions in air showers and
the differences between mass fraction derived using the three
``post-LHC'' models do not necessarily bracket the actual uncertainty
on the fractions. However, barring an onset of new physics in hadronic
interactions at \energy{18.3}, the energy evolution of $\meanXmax$ and
$\sigmaXmax$ are robust indicators of a gradual increase of the
average nuclear mass of cosmic rays with energy.  Further
model-independent evidence for a mixed mass composition around the
ankle was found in a study of correlations between $\Xmax$ and the
shower size measured with the SD~\cite{Aab:2016htd}.

\begin{figure}[t]
\begin{minipage}{.47\textwidth}
\includegraphics[clip, rviewport=0 0.68 1 1,width=\textwidth]{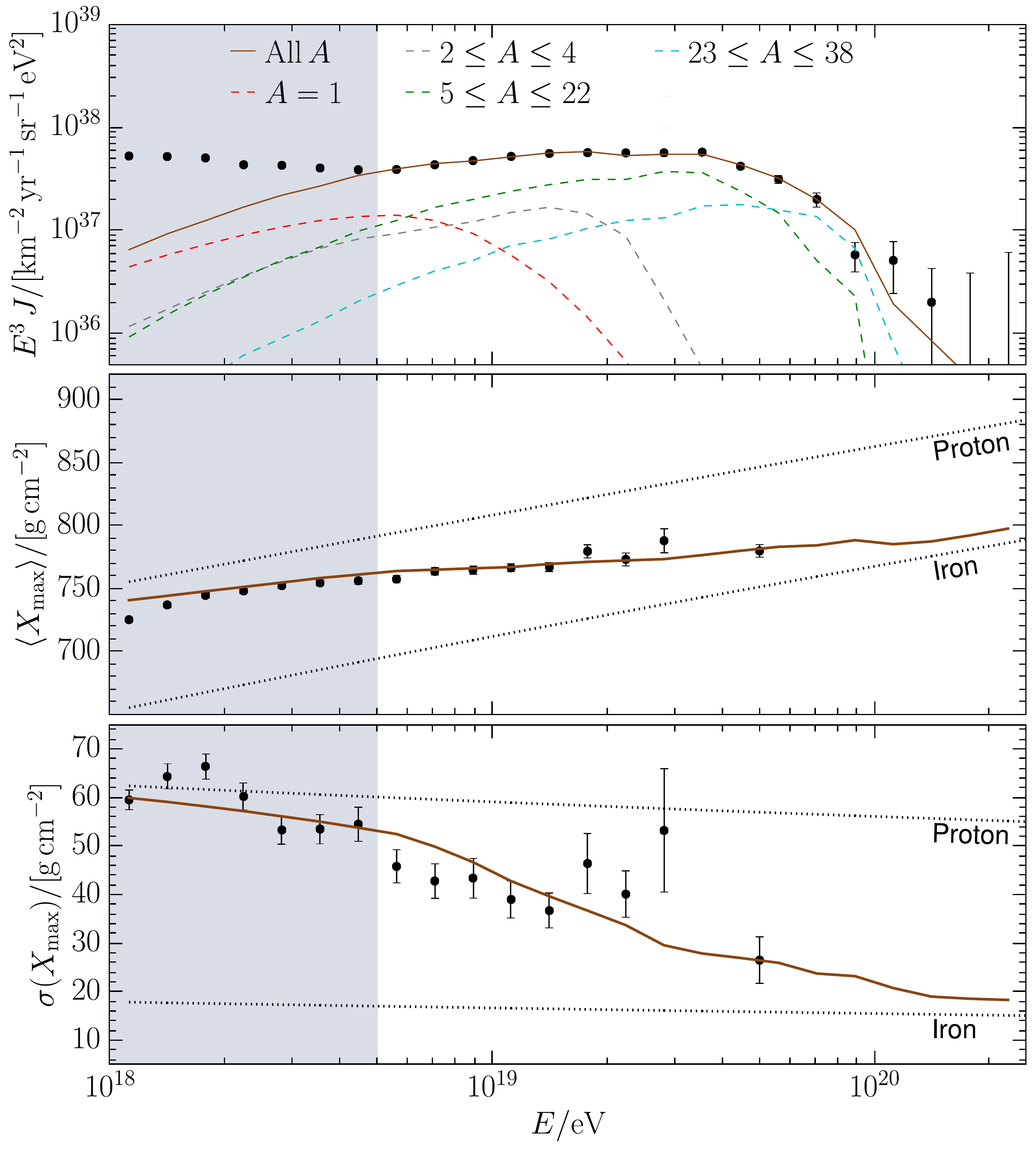}\\
\includegraphics[clip, rviewport=0 0 1 0.069,width=\textwidth]{combinedFit}
\end{minipage}\hfill
\begin{minipage}{.47\textwidth}
\includegraphics[clip, rviewport=0 0 1 0.678,width=\textwidth]{combinedFit}
\end{minipage}
\caption{Mixed composition scenario including
  the effects of a discrete source distribution and the propagation
  of cosmic rays through the extragalactic photon and magnetic fields
  compared to the energy spectrum (left) and $\meanXmax$
   and $\sigmaXmax$ (right). The fit was performed in the
  energy range above the shaded region and the data points are
  measurements from the Pierre Auger Observatory presented at the
  previous ICRC~\cite{ValinoPAO2015, Aab:2014kda}. }
\label{fig:interpretation}
\end{figure}
\section{Interpretation of Mass Composition and Spectrum}
\label{sec:interpret}
For a possible astrophysical interpretation of our results on the mass
composition and energy spectrum, we considered a scenario in which the
sources of UHECRs are of extragalactic origin
and accelerate nuclei in electromagnetic processes with a
rigidity-dependent maximum energy, $E_{\rm max}(Z) = E_{\rm max}({\rm
  p}) / Z$, where $Z$ denotes the charge and $E_{\rm max}({\rm p})$ is
the maximum energy for protons. In a previous study~\cite{Aab:2016zth}
we reported that within this scenario a good description of the shape
of the measured energy spectrum as well as the energy evolution of the
\Xmax distributions can be achieved if the sources accelerate a
primary nuclear mix consisting of p, He, N and Si, if the primary
spectrum follows a power law $\varpropto E^{-\gamma}$ with a spectral index
$\gamma \sim 1$ and if the maximum energy of protons is
about $\energy{18.7}$. In this case, the observed increase in the
average nuclear mass is explained by the disappearance of light nuclei
from the overall composition mix as the energy increases and the flux
suppression is caused by both, energy losses during extragalactic
propagation and maximum energy, of the highest-charge nucleus.

At this conference we presented further investigations of this
astrophysical scenario~\cite{combinedFit}. The homogeneous
distribution of sources assumed in our previous calculations was
replaced by discrete sources distributed according to the model of the
local large-scale structure from~\cite{DolagGST2005} and with a source
density of $10^{-4}/\mathrm{Mpc^{3}}$.
Furthermore, we studied the effects of the extragalactic magnetic field
(EGMF) on the inferred source parameters by tracking the cosmic-ray
trajectories through the EGMF model
of~\cite{BatistaDEKKMSVWW2016} during the propagation from the
sources to Earth.

A fit of the data with this extended model is shown in
Fig.~\ref{fig:interpretation}.  It provides a good overall description
of the data and the goodness of fit is similar to the one achieved
with the simpler model used in our previous study. The energy range
below the ankle is excluded from the fit (as indicated by the shaded
region) to avoid the modeling of possible contributions of Galactic
cosmic rays and/or protons originating from photo-nuclear interactions
in the source environment~\cite{Globus:2015xga, Unger:2015laa}.
Comparing the best-fit parameters of the extended model with our
previous results, we found that the details of the local large-scale
structure of matter are of minor importance for the derived parameters of the
source spectra.  But by including in the calculation the diffusion in
the EGMF, we derive a spectral index of $\gamma \sim 1.6$,
i.e.\ significantly softer than $\gamma \sim 1$ as obtained without magnetic
fields. This softening is caused by the suppression of the cosmic ray
flux at low rigidities when the time it takes cosmic rays to diffuse
to Earth is comparable to the age of the
universe~\cite{MollerachR2013}. Therefore the presence of magnetic fields in the
intergalactic space needs to be taken into account when
interpreting cosmic ray data, especially when the field strength is
relatively strong as assumed in this study.

\section{Searches for Neutrinos and Photons}
\begin{figure}[t]
\centering
\includegraphics[width=0.48\textwidth]{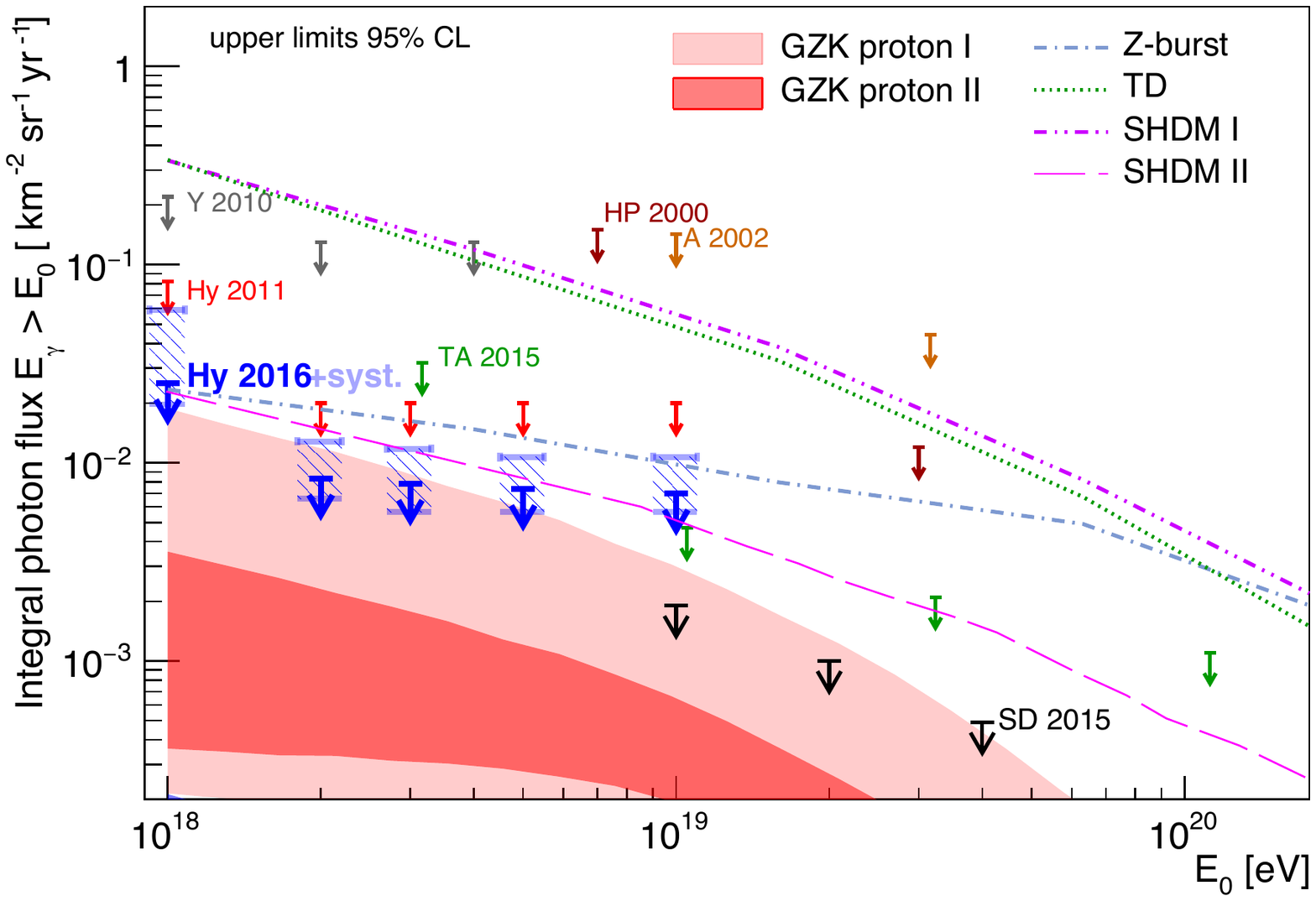}\hfill
\includegraphics[width=0.5\textwidth]{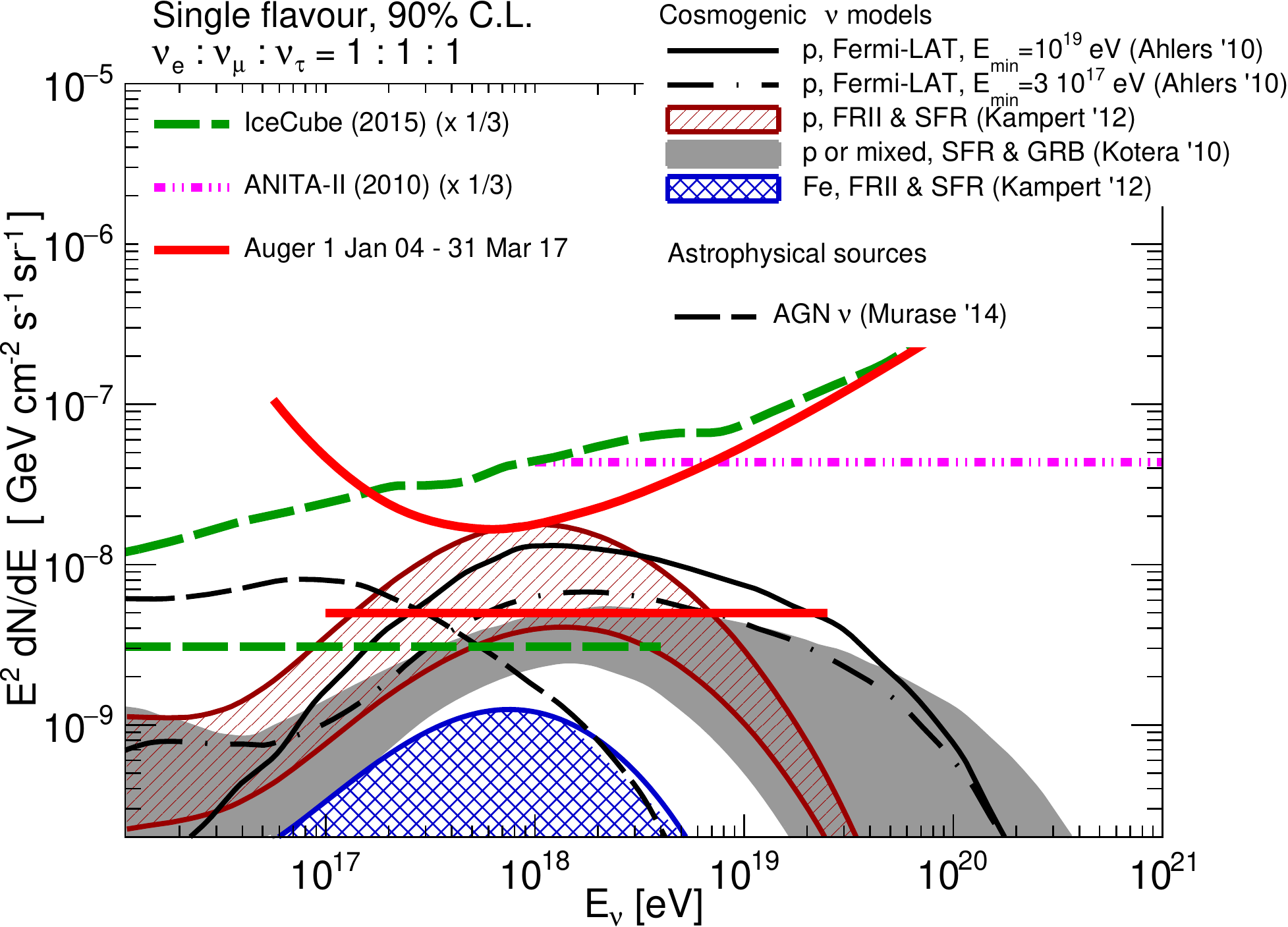}
\caption{Upper limits on the diffuse photon (left) and neutrino  flux (right)
compared to limits from other experiments and model predictions (see~\cite{photons, neutrinos} for references).}
\label{fig:limits}
\end{figure}

Neutrinos and photons are ideal astrophysical messengers for targeted
source studies, since they are not deflected in the Galactic and
extragalactic magnetic fields. In addition, the diffuse flux of
neutrinos and photons carries information about the propagation of
cosmic rays, their mass and the spacial distribution of the cosmic-ray
sources,  These ``cosmogenic'' neutrinos and photons originate from
pions produced in proton-photon interactions with intergalactic
radiation fields and their subsequent decays. A further source of
cosmogenic neutrinos is the beta decay of neutrons produced in the
photo-nuclear interactions of cosmic-ray nuclei during propagation.

The identification of photons with the Pierre Auger Observatory relies
on the fact that photon-induced air showers penetrate deeper in the
atmosphere and produce fewer muons than showers initiated by protons or
nuclei. The best separation power between photons and charged cosmic
rays is achieved by hybrid observations of air showers in which both
the longitudinal development and the particle densities at the ground are
measured~\cite{photons}.

Neutrino-induced air showers are searched for by scanning the data
for upward-going, near-horizontal events (Earth-skimming neutrinos) or
down-going, near-horizontal events with a large electromagnetic
component, i.e.\ a first interaction point that is very deep the
atmosphere~\cite{neutrinos}.

The targeted search for photon point-sources presented at this
conference yielded no evidence for \EeV photon emitters in any of the
studied source classes~\cite{Aab:2016bpi}. No candidate for a
neutrino-induced shower was found and we presented an update of our
previous limit~\cite{Abreu:2012zz} on the neutrino flux from steady
point-sources. The absence of neutrino events around the time of
gravitational wave events produced by binary Black Hole (BH) mergers
reported by the LIGO Collaboration~\cite{gw} allowed us to constrain
the amount of total energy emitted in \EeV neutrinos by black hole
mergers~\cite{Aab:2016ras}.

Results on the diffuse flux of high energy photons and neutrinos are
shown in Fig.~\ref{fig:limits}. As can be seen, our upper limits on
the diffuse flux of photons are the most stringent limits to date and
severely constrain ``top-down'' models in which it is assumed that
UHECRs are the decay products of either super-heavy
dark matter (SHDM), topological defects (TD) or ${\rm Z}^0$ bosons created
in interaction of extremely high energy neutrinos with the relic
neutrino background (Z-burst). With our current sensitivity we probe
photon fractions of about $0.1\%$ and can thus explore the region of photon
fluxes predicted in some astrophysical scenarios for a
proton-dominated mass composition. The updated limits on the diffuse
neutrino-flux show that the sensitivity of the Pierre Auger
Observatory to \EeV neutrinos is comparable to that of dedicated
neutrino experiments. Several predictions for cosmogenic neutrinos that
arise in models explaining the origin of cosmic rays are excluded at
90\%~C.L.,  in particular those that assume proton primaries accelerated
in sources with a strong redshift evolution.

\begin{figure}[t]
\centering
\def\figh{7.2cm}
\includegraphics[clip, rviewport=0.01 0 0.468 1,height=\figh]{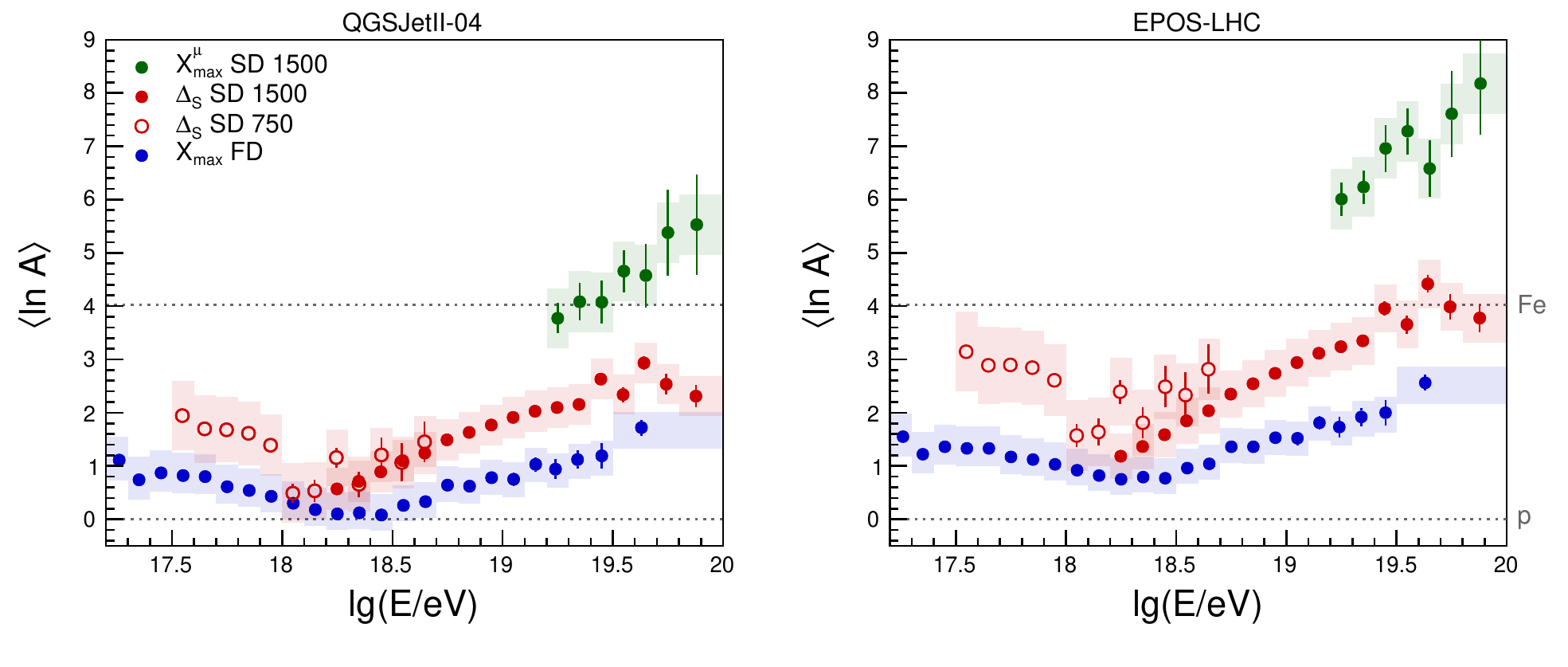}
\includegraphics[clip, rviewport=0.565 0 0.99 1,height=\figh]{lnA}
\caption{Average logarithmic mass, \meanLnA, as a function of energy
  derived from the analyses presented at this conference: \Xmax (FD),
  $\Delta_{\rm S}$ (SD) and \Xmumax (SD).}
\label{fig:lnA}
\end{figure}
\section{Tests of Hadronic Interactions at Ultra-High Energies}
Ultra-high energy air showers provide the opportunity to study hadronic
interactions beyond energies accessible in the laboratory.  In
previous studies we reported the proton-air cross section at $\sqrt{s}
= 57~\TeV$~\cite{Coll.:2012wt}, and a deficit in the predicted
number of muons~\cite{Aab:2014pza} and in the hadronic
component~\cite{Aab:2016hkv} above $\sqrt{s} = 100~\TeV$. Further
inconsistencies in the modeling of hadronic interactions in air
showers were revealed by studies of muon production
depth~\cite{Aab:2014dua} and the azimuthal asymmetry in the risetime
of SD signals~\cite{Aab:2016enk}.

Whereas extrapolations of the $\rm p+p$ cross sections measured at the
LHC agree well with our measurement at high energies, it is more
challenging to predict the muon content of air showers and thus the
signals observed in the SD stations.  The majority of muons in air
showers are created in decays of charged pions when the energy of the
pion is low enough such that its decay length is smaller than its
interaction length. At ultra-high energies it takes several
generations of interactions until the average pion energy is
sufficiently small for pion decay to dominate. As a result, the total
number of muons in an air shower depends on details of hadronic
interactions along a chain of interactions. Hence, even small
differences in the assumed properties of hadronic interactions can
lead to a sizable effect on the predicted muon number when propagated
over several generations of the particle cascade. In contrast, the
longitudinal development, and in particular \Xmax, measured by the FD
is dominated by the first few interactions and is thus less susceptible
to an accumulation of uncertainties of hadronic particle production
over many generations.

The inconsistencies in the modeling of air showers can be quantified
by converting the average properties of the ground signal measured
with the SD to an average logarithmic mass, \meanLnA, to provide a
common scale between these different observables and the \Xmax
measurements from the FD. This is shown in Fig.~\ref{fig:lnA} for
measurements presented at this conference. The average mass, derived
from risetime-related variable $\Delta_{\rm S}$ (obviously {\itshape
  without} the cross-calibration with FD described in
Sec.~\ref{sec:mass}) and from the update of the measurement of the
average muon production depth~\cite{mpd}, is compared to the \meanLnA
estimated from $\meanXmax$. As can be seen, neither of the two models
can satisfactorily describe all three measurements at the same time
and the models completely fail to describe the muon production depths,
unless one considers the possibility of a trans-iron composition at
ultra-high energies.

We conclude that even though recent hadronic interaction models were
tuned to LHC data there is still ample room for further improvements.
A revision of the models to bring the mass estimates from
ground-level measurements in agreement to the ones from \Xmax will not
only make the SD a more reliable tool for mass composition studies,
but also improve the uncertainty of the modeling of \Xmax
itself~\cite{Ostapchenko:2016bir, TanguyHighlight}.

\section{Arrival Directions}
\label{sec:arri}
At this conference the Pierre Auger Coll. announced the
observation of a large-scale anisotropy in the arrival directions of
cosmic rays above \Energy{8}{18}~\cite{dipole, Aab:2017tyv}. Thanks to
additional ${\sim}2.6$ years of data and an optimized event
selection, earlier hints~\cite{ThePierreAuger:2014nja} of a
non-vanishing amplitude in the first harmonic in right ascension
could be confirmed with a significance of $5.2\sigma$.

\begin{figure}[t]
\centering
\begin{overpic}[clip, rviewport= 0.005 0 1 0.99,width=1\textwidth]{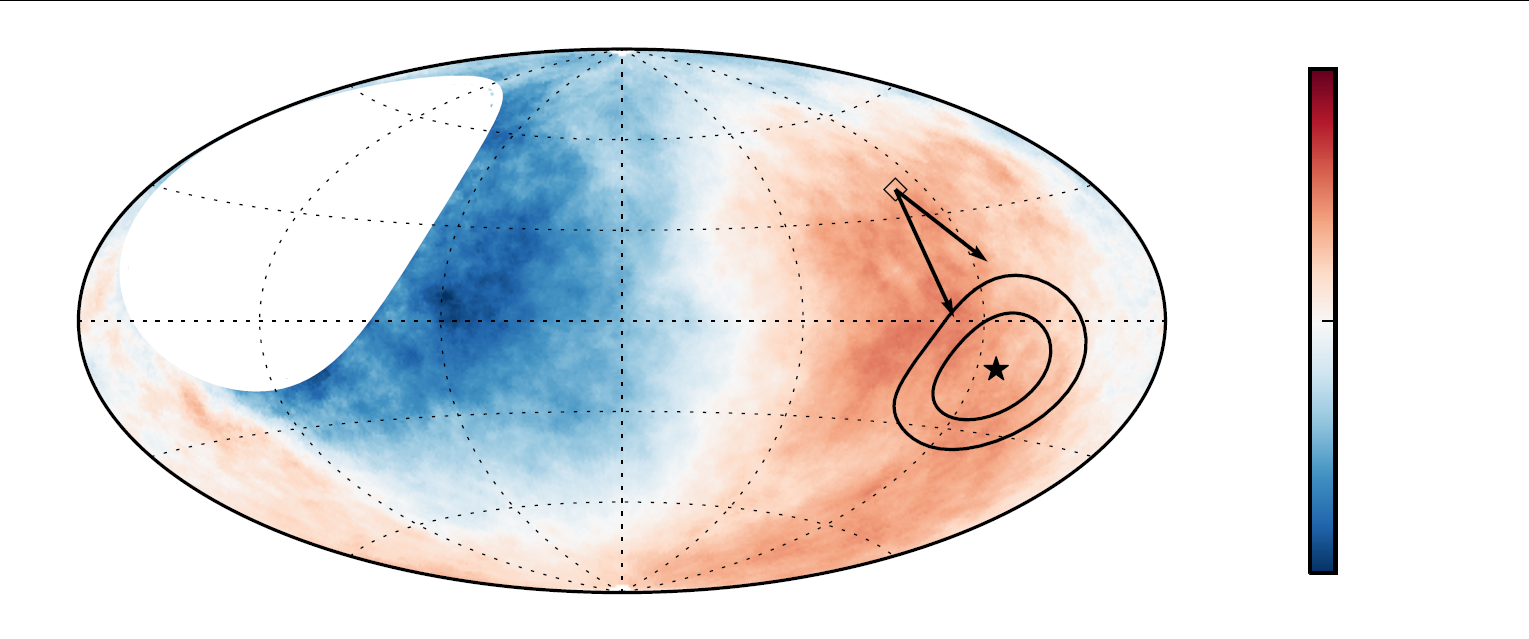}
\put (38,39.5) {\scriptsize $+90^\circ$}
\put (38,1) {\scriptsize $-90^\circ$}
\put (-1,20.4) {\scriptsize $+180^\circ$}
\put (77,20.4) {\scriptsize $-180^\circ$}
\put (88,20.4) {\scriptsize $0.42$}
\put (88,36.8) {\scriptsize $0.46$}
\put (88,4) {\scriptsize $0.38$}
\put (93,10) {\footnotesize \rotatebox{90}{intensity / $\rm (km^2~sr~yr)^{-1}$}}
\put (51.2,30) {\scriptsize \bf {2MRS}}
\put (56.5,21) {\scriptsize \bf 2~EV}
\put (64.5,25.7) {\scriptsize \bf 5~EV}
\end{overpic}
\caption[Flux map]{Map of the intensity in Galactic coordinates for $E \geq 8$\,EeV,
  smoothed by $45^\circ$. The reconstructed dipole direction is
  indicated with a star, the contours represent the 68\% and 95\% CL
  regions. The direction of the dipole in the distribution of galaxies
  from 2MRS is shown by a diamond and the arrows indicate how this
  direction would be modified in a particular model of the Galactic
  magnetic field for two rigidities $E/Z$ compatible with the
  composition shown in Fig.~\ref{fig:comp}.}
\label{fig:fmap}
\end{figure}
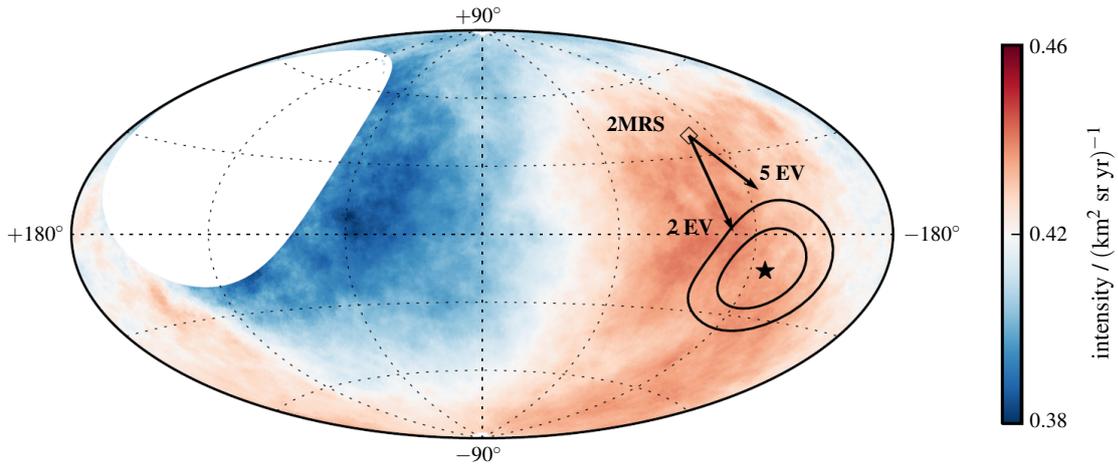

Two energy bins, $4\,\EeV<E<8\,\EeV$ and $E\geq8\,\EeV$, were
monitored since the start of data taking with the Observatory.  The
current measurements of the amplitude of the first harmonic in right
ascension are $0.5_{-0.2}^{+0.6}$\% and $4.7_{-0.7}^{+0.9}$\%,
respectively.  The events in the lower energy bin follow an arrival
distribution consistent with isotropy, but in the higher energy bin a
significant anisotropy was found, with a $p$-value of
$2.6{\times}10^{-8}$ under the isotropic null hypothesis.  A skymap of
the intensity of cosmic rays arriving above \Energy{8}{18} is shown in
Fig.~\ref{fig:fmap}. The data can be well described by a dipole with a
total amplitude of $6.5_{-0.9}^{+1.3}$\%. The reconstructed direction
of the dipole points towards $(\ell,b)=(233^\circ,-13^\circ)$ and is
indicated with a star in Fig.~\ref{fig:fmap}. This direction is about
$125^\circ$ from the Galactic Center, suggesting that the anisotropy
has an extragalactic origin.

A potential cause for a dipole in the arrival directions of cosmic rays
is the cosmological Compton-Getting
effect~\cite{Kachelriess:2006aq}, i.e.\ the anisotropy caused by the
net motion of Earth with respect to the rest frame of UHECR sources. The predicted
amplitude of the  cosmological Compton-Getting
effect is however only 0.6\%, much smaller than the
signal reported here.

Another possibility is that the dipole arises from spatial
inhomogeneities in the distribution of sources or a dominant source
whose image is blurred by the intergalactic and Galactic magnetic
field. For instance, it is known that the distribution of nearby
galaxies, as mapped by the 2MASS redshift survey
(2MRS)~\cite{Huchra:2011ii}, exhibits a dipolar
structure~\cite{Erdogdu:2005wi}. If sources of UHECRs are a subset of
these galaxies, then the arrival direction of cosmic rays at Earth
should follow the same structure.  The dipole of the flux-weighted
distribution of infrared-detected galaxies in the 2MRS catalogue is
shown as an open diamond in Fig.~\ref{fig:fmap}. It is 55$^\circ$ away
from the central direction of the dipole discovered in the arrival
direction of cosmic rays. To illustrate how the Galactic magnetic
field could influence the observed direction of the 2MRS dipole, the
deflected positions of this dipole, as predicted using a particular
model of the Galactic magnetic field~\cite{Jansson:2012pc}, are shown as
arrows in Fig.~\ref{fig:fmap} for two different cosmic-ray rigidities
that are compatible with the composition fractions shown in
Fig.~\ref{fig:comp}. The agreement between the directions of the
UHECR and 2MRS dipoles is improved by adopting these assumptions about the charge
composition and the deflections in the Galactic magnetic field.

The origin of the discovered dipole remains at this point a subject of
speculation. Additional studies of its properties will be performed
soon, for instance the evolution of the dipole with energy and, with
future data from AugerPrime, its dependence on rigidity.  However, it can
already be concluded that the direction of the anisotropy,
pointing away from the Galactic center, provides strong support to the
hypothesis of an extragalactic origin of the highest-energy cosmic
rays.\\

Further studies of the arrival directions of UHECRs were performed at
ultra-high energies and at intermediate angular scales~\cite{arrdir}.
We followed up the two searches with the largest deviations from
isotropy from~\cite{correlation-paper}, now using an exposure of
$9.0\times 10^{4}$\,km$^2$\,sr\,yr, i.e.\ 35\% more than
in our previous study.  We confirm the excess of events (``warm
spot'') above \Energy{5.8}{19} in the direction of the Centaurus~A
radio galaxy within a search radius of $15^{\circ}$ at a post-trial
significance of ${\sim}3.1\sigma$. Furthermore, an excess in the
two-point correlation function between our data and the most
luminous active galactic nuclei (AGNs), detected by
\SwiftBAT~\cite{swift} ($D\leq130$\,Mpc and $L\geq10^{44}$\,erg/s)
above \Energy{6.2}{19} within a search radius of $16^{\circ}$, is found
with a post-trial significance of ${\sim}3.2\sigma$.

\begin{figure}[t]
\centering
\includegraphics[clip,rviewport=0 0 1 0.93,width=0.65\textwidth]{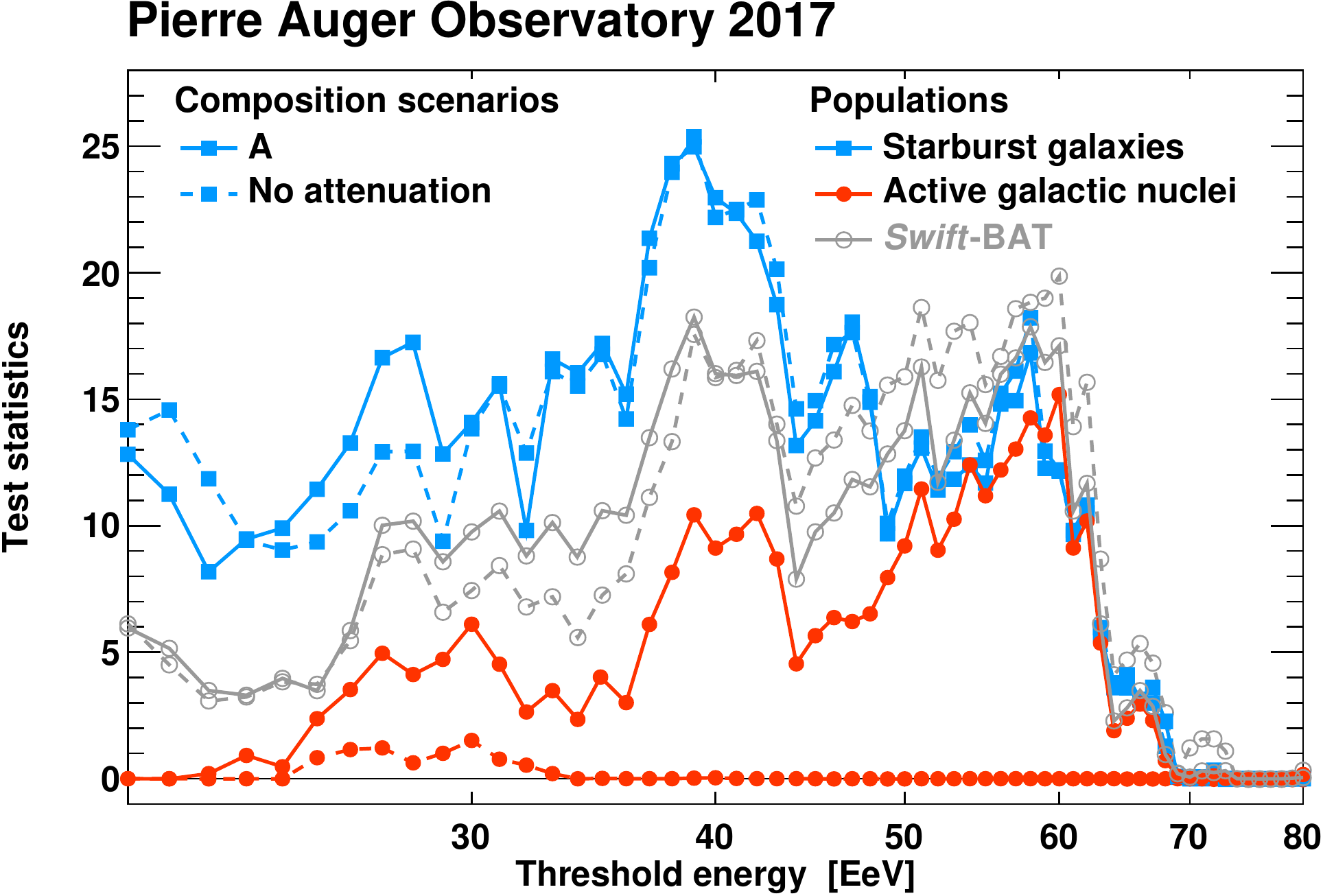}
\caption{Test statistic as a function of the threshold energy for
  starburst galaxies (blue lines), gamma-ray AGNs (red lines) and
  \SwiftBAT AGNs (gray lines). The continuous lines indicate the
  values of the test statistics obtained accounting for attenuation of
  the intensity due to energy losses, while the dotted lines refer to
  the values without any attenuation.}
\label{Escan}
\end{figure}

\begin{figure}[t]
\def\figw{0.47}
\centering
\includegraphics[width=\figw\textwidth]{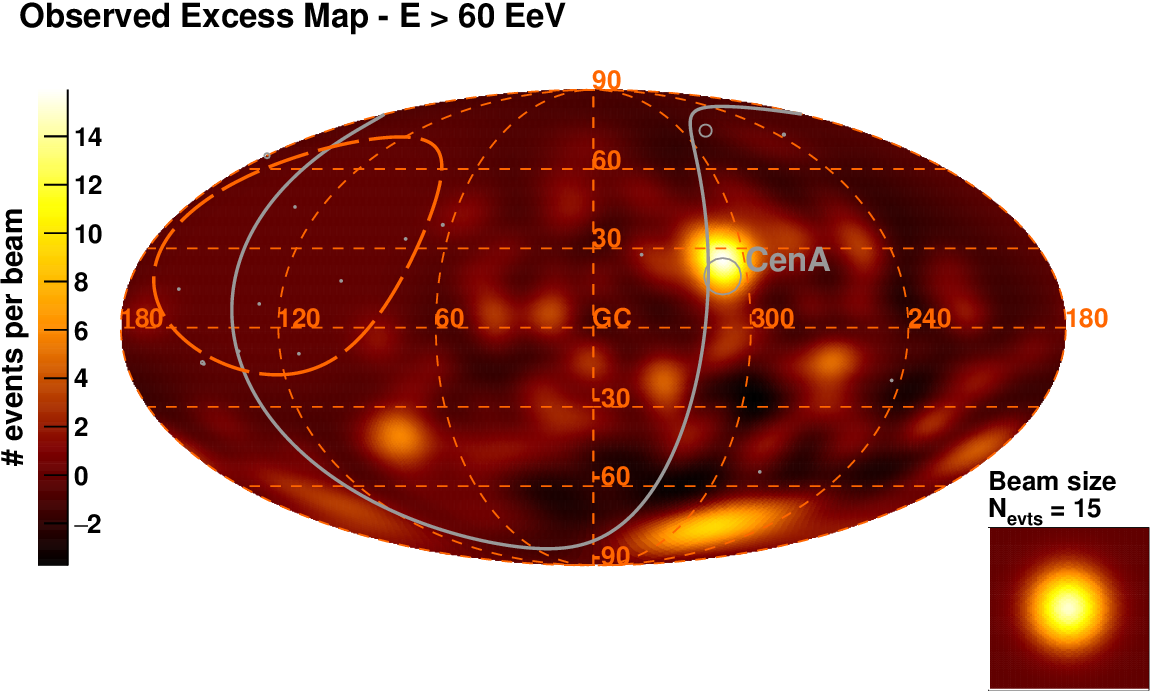}\qquad
\includegraphics[width=\figw\textwidth]{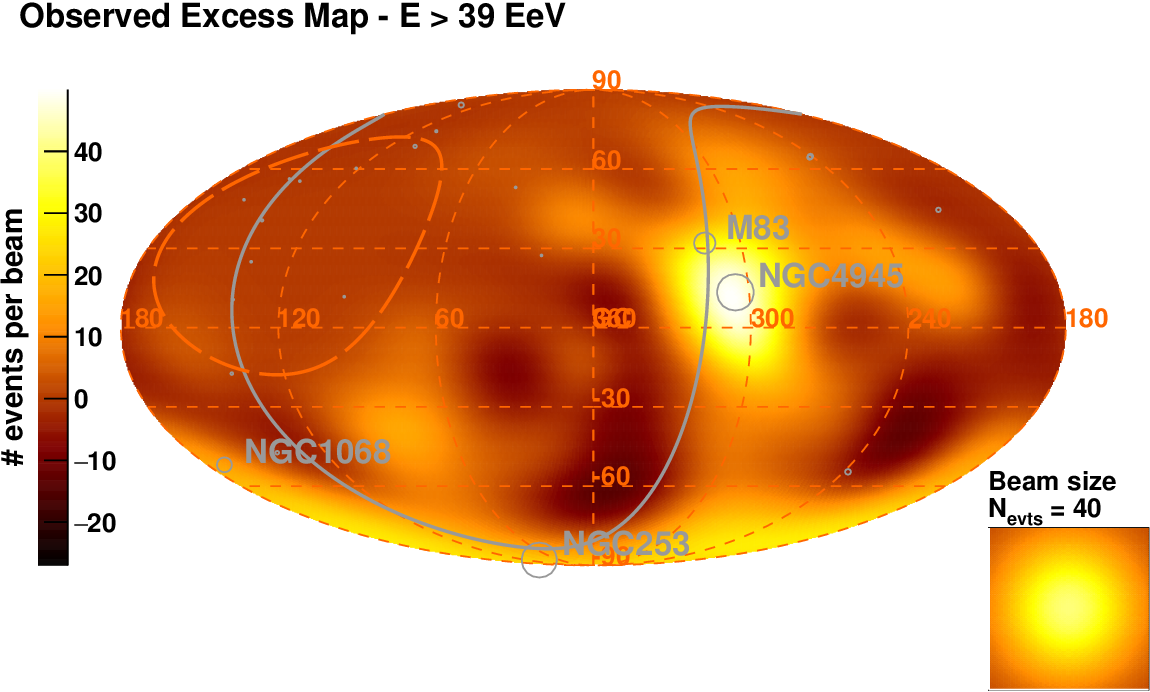}
\\
\includegraphics[width=\figw\textwidth]{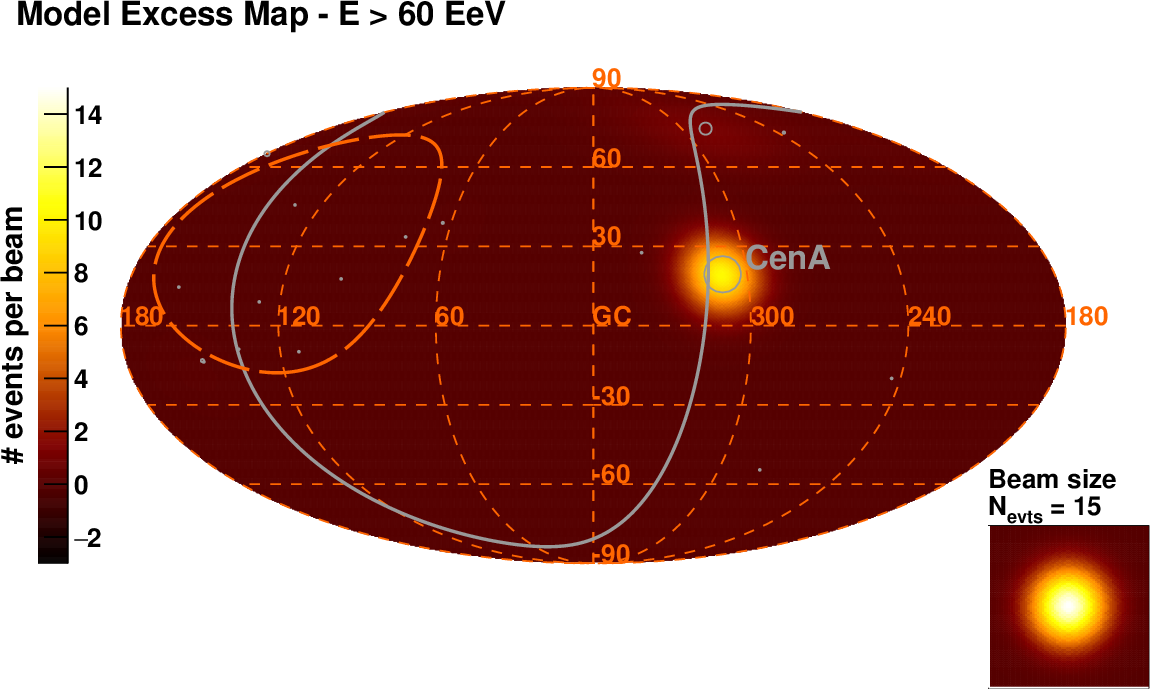}\qquad
\includegraphics[width=\figw\textwidth]{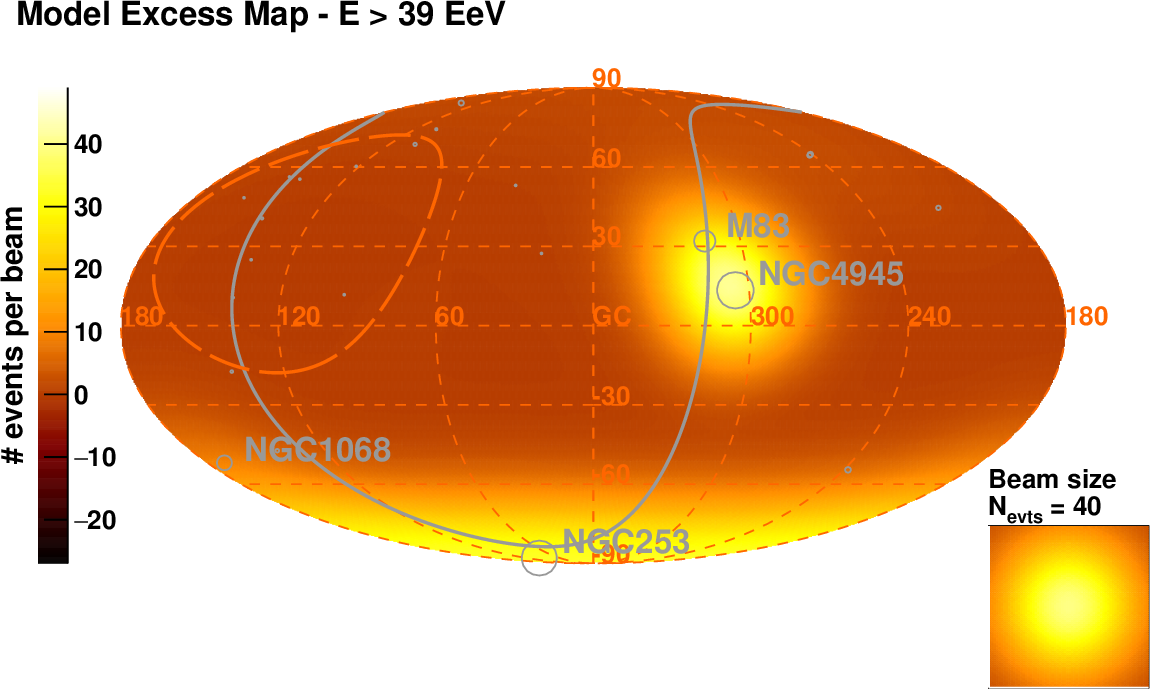}
\caption{Skymaps of the observed (top) and model (bottom) event
  excess with respect to an isotropic background obtained with the
  best-fit parameters for the gamma-ray AGNs (left) and for the
  starburst galaxies (right) in galactic coordinates. The
  supergalactic plane is indicated as a gray line and the limit of the
  field of view of the Observatory is shown as a dashed red
  line. The signal for a particular number of events and the
  respective smearing angle is shown in the insets labeled as ``beam size''.}
\label{SBGmap}
\end{figure}

In a new study we investigated the compatibility of the detected arrival
directions of UHECRs with flux models based on AGNs detected by
\FermiLAT (17 bright nearby AGNs from the 2FHL
Catalog~\cite{2016ApJS..222....5A}) and 23 nearby starburst galaxies
(SBGs)~\cite{HNO}. For the AGNs we assume that the UHECR intensity is
proportional to the integral gamma-ray flux between 50\,\GeV and
2\,\TeV and for the SBGs we used the 1.4\,GHz radio flux as an UHECR
proxy noting that for SBGs detected in gamma rays the radio flux
scales linearly with their gamma-ray
luminosity~\cite{2012ApJ...755..164A}. The attenuation of the
intensity due to energy losses en route to Earth is taken into account
assuming different composition scenarios fitting our
data~\cite{combinedFit}.  The free parameters of this study
are the smearing angle and the fraction of anisotropic cosmic rays originating from the intensity model. A cut on the threshold energy is optimized to maximize the signal and the optimization is penalized for through Monte-Carlo studies.
The evolution
of the test statistic (the likelihood ratio between isotropy
and model for the best-fit isotropic fraction and smearing angle) as a
function of threshold energy is shown in Fig.~\ref{Escan}. As can be
seen, the test statistics is maximized at 60 and 39~\EeV for the AGN and
SBG scenario, respectively. The smearing angle and the anisotropic
fraction corresponding at maximum test statistic are 13$^{\circ}$ and
10\% for the starburst-galaxies and 7$^{\circ}$ and 7\% for the
gamma-ray AGNs.  Comparisons of the skymaps of the smeared data and
SBG/AGN models are shown in Fig.~\ref{SBGmap}.  The post-trial
significance is 2.7$\sigma$ for the gamma-ray AGNs, while for the
starburst galaxies we found a deviation from isotropy at the 4$\sigma$ level.

It is worthwhile noting that all the searches presented here are
\emph{a posteriori} explorations. Numerous studies have been performed
in the past with Auger data within and outside the collaboration. This
makes it difficult to evaluate a proper penalty factor for all the
previous searches. It is however noteworthy that all three
``AGN-type'' models (Centaurus-A only, \SwiftBAT, 2FHL) yield a
post-trial significance at the $3\,\sigma$ level, mainly driven by the
``warm spot'' in the direction of Centaurus-A. The same over-density of
events contributes to the significance of the SBG model, but with a
different interpretation as being caused by the bright starburst
galaxies NGC\,4945 and M83. In addition, the SBG model is able to
describe the mild over-density of events observed around the South
Galactic pole via contributions from NGC\,1068 and NGC\,253. With a
post-trial significance at the $4\,\sigma$ level, the SBG model gives  so
far
the strongest indication for an anisotropy in the arrival directions
of UHECRs at intermediate angular scales found in our data set. Of course correlation does not imply causation and therefore
caution is required in identifying the sources of UHECRs prior to
testing different catalogues and source weights, and a better
understanding of the impact of magnetic deflections in the Galaxy on
the intensity maps of the models.

\section{Outlook: AugerPrime}
The main results derived from the data collected with the Pierre Auger Observatory
so far can be summarized as follows: The energy spectrum of UHECRs
shows a pronounced break at around 5~\EeV (the ankle) and a flux
suppression at around 40~\EeV. Above \energy{17.2} the average mass
decreases, reaching a light composition at \energy{18.3} and increases
again towards ultra-high energies with a possible hint of a change in
the mass evolution above \energy{19.5}. Our non-detection of neutrinos
and photons severely constrains top-down models and limits the
redshift evolution allowed for proton accelerators.  We measured the
proton-air cross section at $\sqrt{s} = 57~\TeV$ and found it in good
agreement with extrapolations from LHC energies, but our SD data
reveals insufficiencies in the modeling of the hadronic component of
air showers. On large angular scales the sky is anisotropic with a
dipolar amplitude of about 7\% at \Energy{8}{18} and there are
indications for anisotropies at intermediate angular scales at
ultra-high energies. These are obviously dramatic advancements of our
understanding of UHECRs, but many open questions still remain. The
\begin{figure}[t]
\def\figh{0.19}
\centering
\includegraphics[clip,rviewport=0 0 1 1,height=\figh\textheight]{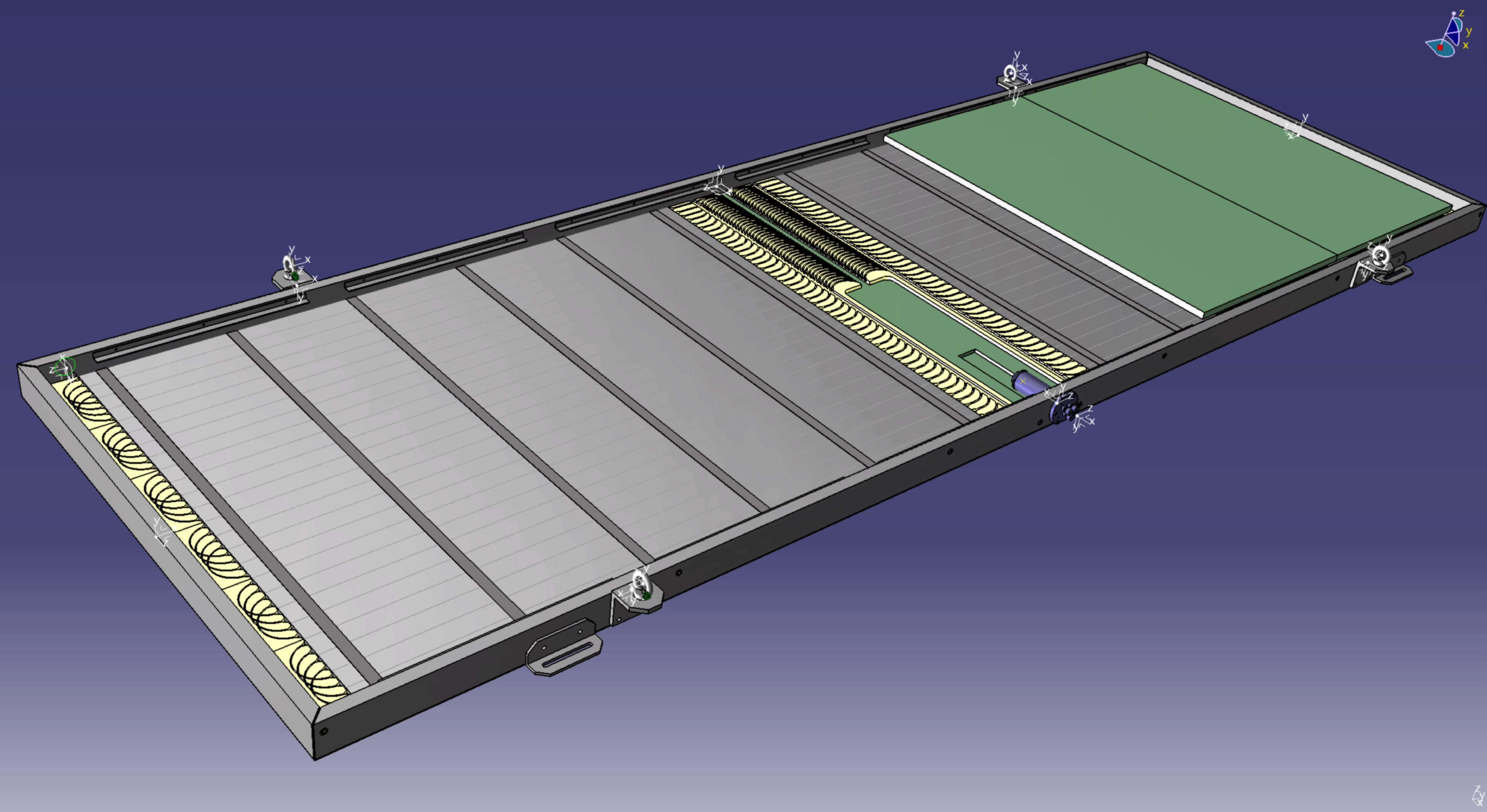}\qquad
\includegraphics[height=\figh\textheight]{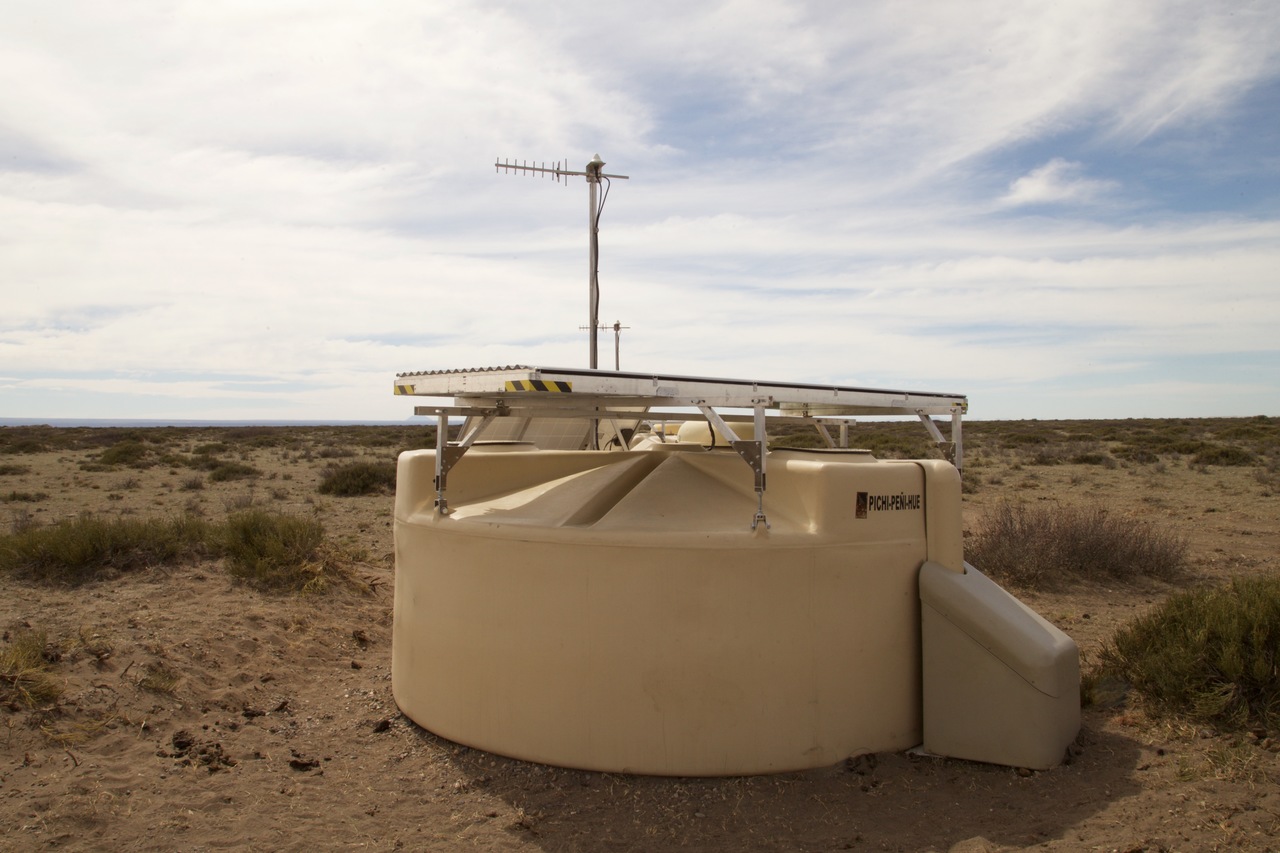}
\caption{{\itshape Left:} Layout of the Surface Scintillator
  Detector (SSD); {\itshape Right:} One station of the AugerPrime
  Engineering Array.}
\label{fig:SSD}
\end{figure}
Accordingly the Pierre Auger Coll. is currently upgrading the Observatory to
address the following important problems:\vspace*{0.2cm}\\
\noindent
{\bf Origin of the flux suppression:}~~Is the flux suppression
caused by energy losses of cosmic rays during the propagation from
their sources to Earth or by the maximum energy of the astrophysical
accelerators? Or maybe a combination of both?\vspace*{0.1cm}\\
\noindent
{\bf Prospects for particle astronomy:}~~What is the fraction of
light elements at ultra-high energies? Is it large enough to perform
charged particle astronomy with tolerable distortions from
Galactic and extragalactic magnetic fields? How do the anisotropies
reported in Sec.~\ref{sec:arri} depend on the particle rigidity?\vspace*{0.1cm}\\
\noindent
{\bf Fundamental physics at ultra-high energies:}~~Are the
inconsistencies between air shower simulations and the surface
detector data due to fundamental shortcomings in our understanding of
hadronic multiparticle production? Can we constrain new physics
phenomena, such as Lorentz invariance violation or extra dimensions at
energies beyond those accessible at human-made accelerators?\\

\vspace*{-0.3cm}
The main focus of the upgrade of the Observatory, named ``AugerPrime'', is
to provide a high-statistics sample of events with mass information at
ultra-high energies which will be achieved by increasing the amount
of information extracted from air shower with the SD~\cite{Aab:2016vlz,
  AugerPrimeOverview}.  Each of the surface detector stations will be
equipped with an additional 4~m$^2$ Surface Scintillator Detector (SSDs)
installed on top of the existing Water-Cherenkov
Detectors (WCDs)~\cite{AugerPrimeSCNT}. The layout of the upgraded
surface detector stations is shown in Fig.~\ref{fig:SSD}. The WCD and
SSD have a different response to the electromagnetic and muonic
component of an air shower, and therefore the combined measurement
allows to disentangle these two components and to provide an estimate
of both, mass and energy of the shower, on an event-by-event
basis. The detectors will be read out by new electronics with a faster
and more accurate sampling of the
signal~\cite{AugerPrimeElectronics}. Together with an extra small
photomultiplier installed in each WCD the current dynamic range will
be extended to more than 32 times the largest signals currently
measured~\cite{AugerPrimeSD}. This setup is complemented by an
Underground Muon Detector~\cite{AMIGA} in the current SD\,750 array
that will be used for the verification and fine-tuning of the methods
used to extract muon information from the SSD and WCD measurements. Furthermore,
we are testing an extension of the operational mode of the FD into periods with a higher night sky background to increase its current duty cycle by 50\%.

At this conference we presented the first results from the AugerPrime
Engineering Array~\cite{AugerPrimeResults} consisting of 12 AugerPrime
detector stations that have been in operation since 2016. With this
setup we have verified the basic functionality of the detector
design, the linearity of the scintillator signal, the calibration
procedures and operational stability. The physics potential of the
upgraded stations is demonstrated by the first measurements of the lateral
distribution of signals with the WCDs and SSDs. The construction of
AugerPrime is expected to be finished by 2019 and it will take data
until 2025.  In this period, the number of events collected will be
comparable with the statistics recorded up to now with the existing
Observatory, but with the advantage of a mass estimate for each event
and thus additional information to elucidate the origin of cosmic rays
at ultra-high energies.

\setlength{\columnsep}{0.2cm}
\begin{multicols}{2}

\end{multicols}
\end{document}